\documentclass[a4paper,12pt]{article}

\pdfoutput=1

\usepackage{xcolor}
\usepackage{appendix}
\usepackage{hyperref}

\usepackage{epsfig}
\usepackage{amssymb}
\usepackage{amsfonts}
\usepackage{amsmath}
\usepackage{euscript}
\usepackage{verbatim}
\usepackage{latexsym}
\usepackage{graphicx}
\usepackage{slashed}
\usepackage[utf8]{inputenc}
\usepackage{graphicx}	
\usepackage{subcaption}
\usepackage{wrapfig}
\usepackage[T1]{fontenc}
\usepackage{mathtools}
\usepackage{caption}
\usepackage{float}
\usepackage{subfloat}
\usepackage{calligra}
\usepackage{braket}
\usepackage{caption}
\usepackage{url}

\usepackage{tikz}
\usetikzlibrary{shapes.geometric, arrows,patterns,snakes}
\tikzstyle{ellip} = [ellipse, minimum width=3cm, minimum height=1cm,text centered, draw=black]

\newskip\humongous \humongous=0pt plus 1000pt minus 1000pt

\newif\ifdtup

\usepackage{xparse}
\ExplSyntaxOn
\NewDocumentCommand{\mref}{m}{\quinn_mref:n {#1}}
\seq_new:N \l_quinn_mref_seq
\cs_new:Npn \quinn_mref:n #1
 {
  \seq_set_split:Nnn \l_quinn_mref_seq { , } { #1 }
  \seq_pop_right:NN \l_quinn_mref_seq \l_tmpa_tl
  ( 
  \seq_map_inline:Nn \l_quinn_mref_seq
    { \ref{##1},\nobreakspace } 
  \exp_args:NV \ref \l_tmpa_tl 
  ) 
 }
\ExplSyntaxOff

\catcode`\@=11

\@addtoreset{equation}{section}

\def\@normalsize{\@setsize\normalsize{15pt}\xiipt\@xiipt
\abovedisplayskip 14pt plus3pt minus3pt%
\belowdisplayskip \abovedisplayskip
\abovedisplayshortskip \z@ plus3pt%
\belowdisplayshortskip 7pt plus3.5pt minus0pt}

\def\small{\@setsize\small{13.6pt}\xipt\@xipt
\abovedisplayskip 13pt plus3pt minus3pt%
\belowdisplayskip \abovedisplayskip
\abovedisplayshortskip \z@ plus3pt%
\belowdisplayshortskip 7pt plus3.5pt minus0pt
\def\@listi{\parsep 4.5pt plus 2pt minus 1pt
     \itemsep \parsep
     \topsep 9pt plus 3pt minus 3pt}}

\relax

\catcode`@=12

\catcode`\@=11

\def\section{\@startsection{section}{1}{\z@}{3.5ex plus 1ex minus
   .2ex}{2.3ex plus .2ex}{\large\bf}}

\def\SymBoxes#1#2#3#4{\newdimen\un@t \un@t#3%
\raisebox{#1}{\rule{#2\un@t}{#4}\hskip-#2\un@t
\@tempdimb\un@t \advance\@tempdimb by-#4\@tempcntb#2\relax%
\@whilenum{\@tempcntb>0}\do{
\rule{#4}{\un@t}\hskip\@tempdimb \advance\@tempcntb by\m@ne}%
\hskip-#2\un@t \rule[\un@t]{#2\un@t}{#4}%
\rule[\un@t]{#4}{#4}\hskip-#4
\rule{#4}{\un@t}}\hskip-#4}                

\begin{document}

\newcommand{\beq}{\begin{equation}}
\newcommand{\eeq}{\end{equation}}
\newcommand{\bea}{\begin{eqnarray}}
\newcommand{\eea}{\end{eqnarray}}
\newcommand{\beas}{\begin{eqnarray*}}
\newcommand{\eeas}{\end{eqnarray*}}
\newcommand{\defi}{\stackrel{\rm def}{=}}
\newcommand{\non}{\nonumber}
\newcommand{\bquo}{\begin{quote}}
\newcommand{\enqu}{\end{quote}}
\renewcommand{\(}{\begin{equation}}
\renewcommand{\)}{\end{equation}}
\def \eqn#1#2{\begin{equation}#2\label{#1}\end{equation}}
\newcommand{\sech}{\text{sech }}

\def\e{\epsilon}
\def\IZ{{\mathbb Z}}
\def\IR{{\mathbb R}}
\def\IC{{\mathbb C}}
\def\IQ{{\mathbb Q}}
\def\de{\partial}
\def\Tr{ \hbox{\rm Tr}}
\def\H{ \hbox{\rm H}}
\def\HE{ \hbox{$\rm H^{even}$}}
\def\HO{ \hbox{$\rm H^{odd}$}}
\def\K{ \hbox{\rm K}}
\def\Im{ \hbox{\rm Im}}
\def\Ker{ \hbox{\rm Ker}}
\def\const{\hbox {\rm const.}}
\def\o{\over}
\def\im{\hbox{\rm Im}}
\def\re{\hbox{\rm Re}}
\def\bra{\langle}\def\ket{\rangle}
\def\Arg{\hbox {\rm Arg}}
\def\Re{\hbox {\rm Re}}
\def\Im{\hbox {\rm Im}}
\def\exo{\hbox {\rm exp}}
\def\diag{\hbox{\rm diag}}
\def\longvert{{\rule[-2mm]{0.1mm}{7mm}}\,}
\def\a{\alpha}
\def\dag{{}^{\dagger}}
\def\tq{{\widetilde q}}
\def\p{{}^{\prime}}
\def\W{W}
\def\N{{\cal N}}
\def\hsp{,\hspace{.7cm}}

\def\br{\nonumber\\}
\def\IZ{{\mathbb Z}}
\def\IR{{\mathbb R}}
\def\IC{{\mathbb C}}
\def\IQ{{\mathbb Q}}
\def\IP{{\mathbb P}}
\def \eqn#1#2{\begin{equation}#2\label{#1}\end{equation}}

\def\){\right)}
\def\({\left( }
\def\]{\right] }
\def\[{\left[ }

\newcommand{\C}{\ensuremath{\mathbb C}}
\newcommand{\Z}{\ensuremath{\mathbb Z}}
\newcommand{\R}{\ensuremath{\mathbb R}}
\newcommand{\rp}{\ensuremath{\mathbb {RP}}}
\newcommand{\cp}{\ensuremath{\mathbb {CP}}}
\newcommand{\vac}{\ensuremath{|0\rangle}}
\newcommand{\vact}{\ensuremath{|00\rangle}                    }
\newcommand{\oc}{\ensuremath{\overline{c}}}
\newcommand{\lag}{\langle}
\newcommand{\rag}{\rangle}
\newcommand{\psiin}{\psi_{0}}
\newcommand{\phiin}{\phi_{1}}
\newcommand{\hin}{h_{0}}
\newcommand{\rh}{r_{h}}
\newcommand{\rb}{r_{b}}
\newcommand{\psibnd}{\psi_{0}^{b}}
\newcommand{\psibndp}{\psi_{1}^{b}}
\newcommand{\phibnd}{\phi_{0}^{b}}
\newcommand{\phibndp}{\phi_{1}^{b}}
\newcommand{\gbnd}{g_{0}^{b}}
\newcommand{\hbnd}{h_{0}^{b}}
\newcommand{\zh}{z_{h}}
\newcommand{\zb}{z_{b}}
\newcommand{\man}{\mathcal{M}}
\newcommand{\ee}{\mathbf{e}}
\newcommand{\real}{\text{Re}}
\newcommand{\imag}{\text{Im}}
\newcommand{\caln}{\mathcal{N}}
\newcommand{\f}{\phi}
\newcommand{\m}{\mu}
\newcommand{\n}{\nu}
\newcommand{\z}{\zeta}
\newcommand{\F}{\Phi}
\newcommand{\g}{\gamma}
\newcommand{\G}{\Gamma}
\newcommand{\ci}{\mathcal{I}}
\newcommand{\cj}{\mathcal{J}}
\newcommand{\ck}{\mathcal{K}}
\newcommand{\bphi}{b^{\Phi}}
\newcommand{\cor}{c_{\mathcal{R}}^{\Omega}}
\newcommand{\boi}{b_{\mathcal{I}}^{\Omega}}
\newcommand{\calg}{\mathcal{G}}
\newcommand{\cald}{\mathcal{D}}
\newcommand{\dslash}{\partial\hspace{-.25cm}/}
\newcommand{\delslash}{\nabla\hspace{-.3cm}/}
\newcommand{\zetabar}{\overline{\zeta}}
\newcommand{\calw}{\mathcal{W}}
\newcommand{\deps}{\delta_{\epsilon}}
\newcommand{\dtn}{_{10}}
\newcommand{\calv}{\mathcal{V}}
\newcommand{\calo}{\mathcal{O}}
\newcommand{\zz}{$ SU(2)\times SU(2)\times\IZ_2 \ $}
\newcommand{\lr}[1]{\big\langle #1 \big\rangle}
\newcommand{\exval}[1]{\big\langle \mathcal{O}_{#1}\big\rangle}
\newcommand{\exvalb}[2]{\big\langle \overline{\mathcal{O}}^{#1}_{#2} \big\rangle}
\newcommand{\ol}[1]{\hat{#1}}
\newcommand{\dep}{\delta_{\epsilon}}
\newcommand{\des}{\delta_{\sigma}}
\newcommand{\depm}{\delta_{\epsilon^{-}}}
\newcommand{\hzd}[1]{\hat{\zeta}_{\ol#1}}
\newcommand{\qft}{_{\text{\tiny QFT}}}
\newcommand{\cals}{\mathcal{S}}
\newcommand{\la}{\log a z\,}
\newcommand{\gs}{g_{s}}
\newcommand{\hks}{h_{\tx{KS}}}
\newcommand{\limo}{\lim\limits_{z\rightarrow 0}}
\newcommand{\hzdb}[1]{\overline{\hat{\zeta}}_{\ol#1}}
\newcommand{\tx}{\text}

\begin{titlepage}
\begin{flushright}
CHEP XXXXX
\end{flushright}
\bigskip
\def\thefootnote{\fnsymbol{footnote}}

\begin{center}
{\Large
{\bf On the KKLT Goldstino \\ \vspace{0.1in} 
}
}
\end{center}

\bigskip
\begin{center}
{\large  Chethan KRISHNAN$^a$\footnote{\texttt{chethan.krishnan@gmail.com}}, \ \ Himanshu RAJ$^{b,c}$\footnote{\texttt{hraj@sissa.it}}, \\ 
\vspace{0.2cm}
\& P. N. Bala SUBRAMANIAN$^a$\footnote{\texttt{pnbalasubramanian@gmail.com}}}
\vspace{0.1in}

\end{center}

\renewcommand{\thefootnote}{\arabic{footnote}}

\begin{center}

$^a$ {Center for High Energy Physics,\\
Indian Institute of Science, Bangalore 560012, India}\\

\bigskip

$^b$ {Mani L. Bhaumik Institute for Theoretical Physics,}\\
{Department of Physics and Astronomy, }\\
{University of California, Los Angeles, CA 90095, USA}

\bigskip

$^c$ {SISSA and INFN, \\
Via Bonomea 265; I 34136 Trieste, Italy}\\

\end{center}

\noindent
\begin{center} {\bf Abstract} \end{center}
We construct general asymptotically Klebanov-Strassler solutions of a five dimensional $SU(2) \times SU(2) \times \IZ_2\times \IZ_{2R}$ truncation of IIB supergravity on $ T^{1,1} $, that break supersymmetry. This generalizes results in the literature for the $SU(2) \times SU(2) \times \Z_2\times U(1)_R$ case, to a truncation that is general enough to capture the deformation of the conifold in the IR. We observe that there are only two SUSY-breaking modes even in this generalized set up, and by holographically computing Ward identities, we confirm that only one of them corresponds to spontaneous breaking: this is the mode triggered by smeared anti-D3 branes at the tip of the warped throat. Along the way, we address some aspects of the holographic computation of one-point functions of marginal and relevant operators in the cascading gauge theory. Our results strengthen the evidence that {\it if} the KKLT construction is meta-stable, it is indeed a spontaneously SUSY-broken (and therefore bona fide) vacuum of string theory.

\vspace{1.6 cm}
\vfill

\end{titlepage}

\setcounter{footnote}{0}

\section{Introduction and Conclusion}

Controllably breaking supersymmetry  (SUSY) in supersymmetric theories is generally a difficult problem. This raises a challenge for (super)string theory, because the real world is non-supersymmetric and has a positive cosmological constant, which means that for string theory to be phenomenologically viable \cite{Willy}, it needs to admit (likely meta-stable\footnote{It is possible that metastable SUSY-breaking vacua \cite{ISS,Argurio:2006ny,Argurio:2007qk} are generic in supersymmetric theories with complicated potentials, even if they have supersymmetric vacua elsewhere in the potential landscape. Such vacua have also been found to be cosmologically viable \cite{WillyVadimCK}.}) de Sitter vacua. 
 
The first example of such a de Sitter vacuum in string theory was constructed by KKLT \cite{KKLT}. They did this by considering a fully moduli stabilized warped AdS compactification \cite{GiddingsKP} and proceeding to place a small number $p$ of anti-D3 branes in this warped geometry\footnote{This whole program relies on the existence of flux vacua. Our work does not have much to say directly about this point: our concern is with the nature of SUSY-breaking in them {\em assuming} they exist. This assumption is implicit in all of these works, but see the recent paper \cite{Sethi} which challenges the conventional wisdom.}. The idea is that this breaks supersymmetry and produces positive vacuum energy (which is hierarchically small because of the warping in the geometry) while having a fully stabilized compactification. 

In concrete discussions of KKLT, it is often useful to think of a non-compact Calabi-Yau geometry called the conifold, instead of a fully stabilized compact space. In this non-compact setting, one adds anti-D3 branes \cite{KPV} to the tip of the so-called warped deformed conifold geometry, which is known to be holographically dual to an ${\cal N}=1$ non-conformal gauge theory called the Klebanov-Strassler (KS) cascading\footnote{The duality cascade in turn can be understood \cite{Jarah} via the mechanism suggested in \cite{KPV}.} gauge theory \cite{KS}. The advantage of considering this set up is threefold. Firstly, it enables us to modularize the problem: one can address questions that are not tied to the technicalities of stabilizing the compactification in this more relaxed context, and then hope to ``attach'' the result to a fully stabilized compact Calabi-Yau. The fact that the conifold is an example of a generic Calabi-Yau singularity \cite{KW} makes this approach plausible. Secondly, the duality between the warped deformed conifold and the cascading gauge theory enables us to use powerful holographic techniques to address various bulk questions in the geometry. Indeed, this will be our primary strategy in this paper. Thirdly, the Klebanov-Strassler theory gives us a concrete setting where we can do explicit calculations, but whose results are expected to have generic significance. 

For this approach to be of any use however, one needs to make sure that when one adds anti-D3 branes at the tip of the throat, the resulting bulk solution should be interpretable as a state in the dual cascading gauge theory\footnote{Note that it is not easy to determine from the bulk side alone if the cascading geometry with and without anti-D3 branes belongs to the same theory. As far as the supergravity is considered, anti-D3 brane sources are like a boundary condition in the IR, and are in some sense arbitrary. Whether they really belong to the spectrum of the gravity theory depends on the UV completion of the supergravity into string theory, and is something which we do not know well because we do not have full control on Klebanov-Strassler as a string background. What holography and the dual cascading theory does here, is to give us a non-perturbative definition of the theory so that we can in principle ask whether certain states belong to its spectrum. 
}. In particular, since the anti-branes break bulk supersymmetry, the corresponding state in the dual theory should be one where SUSY is spontaneously broken, which means that it should be characterized by a goldstino mode. Such a mode was indeed identified  in \cite{DKM} and later in \cite{Bertolini} within the context of a certain five dimensional $SU(2) \times SU(2) \times \Z_2\times U(1)_R$ truncation of IIB supergravity on $T^{1,1}$ using the holographic renormalization technology developed by \cite{Aharony}. 

In this paper, our goal is to extend these results to an $SU(2) \times SU(2) \times \IZ_2 \times \IZ_{2R}$ truncation by including supergravity fields which are not neutral under the R-symmetry: the previous constructions \cite{DKM,Bertolini} were working with Klebanov-Tseytlin (KT) \cite{KT} asymptotics, whereas we will deal with the full Klebanov-Strassler.  Klebanov-Tseytlin geometry is singular in the IR and cannot incorporate the deformation of the conifold, while Klebanov-Strassler is a fully regular solution. The (implicit or explicit) hope of the calculations in \cite{DKM,Bertolini} was that since the deformation parameter is a supersymmetric perturbation, it is unlikely to destroy the claims about the SUSY-breaking perturbations. But it must be borne in mind that to discuss this question adequately, one must work with a fully consistent truncation that {\em allows} the deformation in the first place, and see whether (a) such a truncation allows for more SUSY-breaking parameters in the UV asymptotic solution\footnote{That is, a truncation that captures the conifold deformation parameter in the IR has more fields, and might allow more SUSY-breaking parameters in the UV. This cannot be settled merely by looking at the $U(1)$ truncation, one needs at least the $\IZ_2$ truncation.}, and (b) whether the holographic Ward identities \cite{Bertolini,Argurio} get modified in any substantive way. We will answer both these questions in the negative in this paper, by working with the $ SU(2) \times SU(2) \times \IZ_2 \times \IZ_{2R} $ truncation. 

The price we pay for working with a more realistic truncation is that there are extra fields in the system which make the problem more complicated. More conceptually, we will see that the extra fields that we turn on correspond to relevant sources, and that the mixing\footnote{Supergravity fields are not naturally diagonal in the field theory scaling dimension, and so we need to work with appropriate combinations of fields.} of fields that they cause on the supergravity side needs to be suitably taken care of. In the $U(1)$ case, only marginal sources were present and their mixing was dealt with \cite{Bertolini} by defining composite supergravity fields which were diagonal in the scaling dimensions. However, this construction is not always unique, and in the case of relevant sources, we find it more convenient to deal with the leading fall-offs of the would-be composite sources directly. We will see that this information is sufficient to compute the one- and two-point functions required for a holographic calculation of the Ward identities.  

In section \ref{sec:2}, we will review the details of our $\IZ_2$ truncation and the resulting 5d effective supergravity action, starting from the 10d type IIB theory. Then in section \ref{ktsection}, we proceed to describe the Klebanov-Tseytlin and Klebanov-Strassler backgrounds in a 5d language and compare their UV asymptotics. In section \ref{sec:4}, we obtain and present the most general SUSY-breaking solution that asymptotes to the Klebanov-Strassler solution perturbatively in the UV. We show that despite the presence of extra supergravity fields w.r.t. the $U(1)$ truncation, here also there are only two such SUSY-breaking parameters. There are new SUSY-preserving parameters (apart from the conifold deformation parameter) that show up in the solution which we safely ignore since they do not contribute to SUSY breaking dynamics. In section \ref{sec:5} we give a holographic derivation of SUSY and trace Ward identities. We begin by setting up the gauge/gravity dictionary. We identify the holographic sources for dual operators (in particular, sources for marginal and relevant operators). This leads to some subtleties because (as we previously mentioned) the supergravity fields are not automatically diagonal to the field theory operators, so we need to consider appropriate combinations of them. Once these sources and their supersymmetric partners are identified in a useful form, we proceed to derive the SUSY Ward identities. We also derive the identities for the Weyl and super-Weyl transformations for completeness (and because we can). Since we are doing these calculations holographically we will be working with the local supersymmetries and diffeomorphisms of the bulk supergravity theory and derive the Ward identities by demanding that the variation of the renormalized on-shell action under these transformations is zero. To do this, we will need the transformations of the sources, which we compute following \cite{Bertolini}. Finally, in Section \ref{sec:6}, we conclude by checking these identities on the vacua dual to the SUSY-breaking solution found in Section \ref{sec:4} by explicitly calculating one-point functions that characterizes the scale of SUSY breaking. The results we find are consistent with the expectations of \cite{DKM} which were presented in the context of the $U(1)$ truncation. Since the deformation of the conifold is a SUSY-preserving parameter, it is reasonable that our results are consistent with those of \cite{DKM}. It is somewhat remarkable that even in this generalized setup, there are no more SUSY-breaking perturbations, on top of the ones found in the $U(1)$ case and that the number of SUSY-breaking parameters in the UV remains two. So in the end, we find that despite the complications involved in the relevant sources, the final Ward identities remain substantively unchanged. In a series of appendices, we give relevant details needed to reproduce the results in the main text. 

\newpage
\section{Dimensional Reduction of Type IIB SUGRA}\label{sec:2}
In this section we give a brief summary of dimensional reduction of type IIB supergravity theory on $T^{1,1}$ which gives rise to a particular ${\cal N}=4$, 5d gauged supergravity. We will truncate this theory to a particular ${\cal N}=2$ subsector that contains the Klebanov-Strassler solution. This truncation will be relevant for the rest of the paper. The interested reader can find more details in \cite{Buchel,Cassani,Bena:2010pr,Halmagyi:2011yd,Liu:2011dw}. 

The type IIB supergravity in the Einstein frame, takes the form
\begin{align}
\begin{split}
S_{10} &= \dfrac{1}{2\kappa_{10}^{2}}\int_{\man_{10}} \( R_{10} - \dfrac{1}{2} \(d\phi\)^2 - \dfrac{1}{2}e^{-\phi} H_{3}^2 - \dfrac{1}{2} e^{\phi} F_{3}^2 - \dfrac{1}{2} e^{2\phi} F_{1}^2- \dfrac{1}{4} F_{5}^2\)\star1\\
&-\dfrac{1}{8\kappa\dtn^{2}}\int_{\man_{10}} (B_{2}\wedge dC_{2} - C_{2}\wedge dB_{2})\wedge dC_{4}~.
\end{split}
\end{align}
The ten dimensional space-time is denoted by $ \man\dtn $. $ \kappa\dtn $ is related to the ten dimensional Newton's constant. The field strengths satisfy the following Bianchi identities
\bea
dF_{1}= 0~,\ \ dH_{3} = 0~, \ \ dF_{3} = H_{3}\wedge F_{1}~, \ \ dF_{5} = H_{3}\wedge F_{3}~.
\eea
The equations of motion of the type IIB supergravity action have to be supplemented with the self-duality condition
\bea
\star\dtn F_{5} = F_{5}~.
\eea
We are interested in reductions of this theory on the coset $ T^{1,1} = (SU(2)\times SU(2))/U(1)$ with the $U(1)$ embedded in the two $SU(2)$'s diagonally. $T^{1,1}$ can be parametrized in terms of polar coordinates $ (\theta_{1},\phi_{1},\theta_{2},\phi_{2},\psi) $, with ranges $ 0\leq \theta_{1,2}< \pi, 0\leq \phi_{1,2}< 2\pi $ and $ 0\leq \psi< 4\pi $ in the following way
\begin{align}\label{oneforms1}
\begin{split}
&\ee^{1} = -\sin\theta_{1} \;d\phi_{1}~, \  \ \ \ee^{2}  = d\theta_{1}~, \\
& \ee^{3} = \cos\psi \sin\theta_{2} \;d\phi_{2} - \sin\psi \;d\theta_{2}~,\\
& \ee^{4} = \sin\psi \sin\theta_{2} \;d\phi_{2} + \cos\psi \;d\theta_{2}~,\\
& \ee^{5} = d\psi + \cos\theta_{1}\; d\phi_{1} +\cos\theta_{2}\;d\phi_{2}~.
\end{split}
\end{align}
The left-invariant 1- and 2-forms are given by \cite{Cassani}
\bea
\begin{split}
\eta = -\dfrac{1}{3}\ee^{5}~, \qquad \Omega = \dfrac{1}{6} (\ee^{1} + i \ee^{2} )\wedge (\ee^{3} - i \ee^{4})~ ,\hspace{0.2in} \\
J = \dfrac{1}{6}(\ee^{1}\wedge \ee^{2} - \ee^{3} \wedge \ee^{4})~,\qquad \Phi = \dfrac{1}{6} (\ee^{1}\wedge \ee^{2} + \ee^{3} \wedge \ee^{4})~.
\label{BasisofT11}
\end{split}
\eea
The dimensional reduction proceeds by factoring the 10d spacetime ${\cal M}_{10}$ into the warped product space ${\cal M}_{10}={\cal M}_5\times_w T^{1,1}$ and expanding out the 10d form fields in the basis of the left invariant one forms \eqref{BasisofT11} (see Section 3.2 of \cite{Cassani}). The 10d scalars $\phi$ and $C_0$ and all the coefficients in this reduction ansatz are taken to be functions of the coordinates on ${\cal M}_5$. Non-trivial cycles of the internal manifold can allow for additional terms in the expansion for the field strengths. This ansatz retains all and only those modes of type IIB supergravity that are invariant under the action of the isometry group of $T^{1,1}$ which is $SU(2)\times SU(2)$ and automatically guarantees the consistency of the reduction. The resulting 5d effective action matches with the structure of 5dimensional $ \mathcal{N}=4 $ gauged supergravity. The field content of the dimensionally reduced theory, along with its type IIB origins, is reproduced from \cite{Cassani} in Table \ref{n=4trunc} below.

\begin{table}[h]
\centering
\begin{tabular}{c|c|c|c|c}
\hline 
IIB fields  & scalars & 1-forms & 2-forms & 5d metric \\ 
\hline 
10d metric & $u,v,w,t,\theta$ & $A$ &   & $g_{\mu\nu}$  \\		 

$\phi$ &  $\phi $ &   &   &   \\ 

$B_2$ & $b^{J},b^{\Phi},b^{\Omega}$ & $b_{1}$ & $b_{2}$ &  \\ 

$C_{0}$ & $C_{0}$ & & & \\

$C_{2}$ & $c^{J},c^{\Phi},c^{\Omega}$ & $c_{1}$ & $c_{2}$ &  \\

$C_{4}$ & $a$ & $a_{1}^{J},a_{1}^{\phi},a_{1}^{\Omega}$ & $a_{2}^{\Omega}$ & \\
\hline 
\end{tabular} 
\caption{5d fields along with their 10d origins.}
\label{n=4trunc}
\end{table}
Apart from these 5d fields, there are also flux terms $p,q$ and $k$ that descends from the expansion of the field strengths with legs along the cohomologically non-trivial cycle $\Phi\wedge\eta$ and the volume form. These parameters appear explicitly in the scalar potential and characterizes the gauging. 

By consistently turning off the following 5d fields, one finds a further truncation to an ${\cal N} = 2$ gauged supergravity
\bea
b_{2} = c_{2} = a_{2}^{\Omega}=a_{1}^{\Omega} = b_{1} = c_{1} =b^{J} = c^{J} = 0~.
\eea
The so-called $\caln = 4$ Betti vector multiplet, consisting of $\{ a_{1}^{\phi}, w, b^{\Phi}, c^{\Phi}, t,\theta \}$, in the original reduction can be viewed as an $\caln = 2$ vector multiplet $\{ a_{1}^{\phi}, w \}$ together with a $\caln = 2$ hypermultiplet $\{b^{\Phi},c^{\Phi}, t,\theta \} $. Setting either of them to zero is a consistent sub-truncation and gives rise to an ${\cal N}=2$ theory. Truncating out the vector multiplet gives rise to an ${\cal N}=2$ gauged supergravity coupled to three hypermultiplets and a vector multiplet which are invariant under a $\mathbb{Z}_2$ symmetry (not to be confused with the $\mathbb{Z}_{2R}$ symmetry associated to the gaugino condensation in the dual field theory).  This symmetry acts in the following way
\begin{itemize}
\item $(\theta_1,\phi_1)~~\leftrightarrow ~~ (\theta_2,\phi_2)$ .
\item Flip the signs of field strengths $H_3$ and $F_3$ (this corresponds to the action of $-\mathbb{I}$ of $SL(2,\mathbb{Z})$ duality group of Type IIB)~.
\end{itemize}
Here $\theta_i$ and $\phi_i$ are the coordinates on $T^{1,1}$. Under the above transformation, the scalar fields $b_J, c_J$ and $w$ flip sign\footnote{This is because in \cite{Cassani}, the 2-form $J$ is invariant and $\(\ee^1\)^2+\(\ee^2\)^2 \leftrightarrow \(\ee^3\)^2+\(\ee^4\)^2$
under the transformation under $(\theta_1,\phi_1) \leftrightarrow (\theta_2,\phi_2)$.}. In the aforementioned sub-truncation these fields are not present. The fields that survive are presented in Table \ref{n=2trunc} below. We can refer to this sector as the $\mathbb{Z}_2$ truncation \cite{Aharony}. On top of this $\IZ_2$ these fields also have an $\IZ_{2R}$ symmetry\footnote{This can be found by looking at the 10d reduction ansatz. The complex 2-form $\Omega$ has an over multiplicative factor of $e^{ - i \psi }$. Since the coordinate $\psi$ is in the range $(0,4 \pi)$, to see the $U(1)_R$ it is convenient to define another coordinate (say) $\sigma = \psi / 2$. In terms of $\sigma$, $\Omega$ has the multiplicative factor $e ^{- 2 i\sigma}$. This means that $\Omega$ has $U(1)$ R-charge $-2$ which implies that $b^\Omega$ and $c^\Omega$ have R-charge $2$. 
This means that the elements of the $U(1)_R$ which leave $\Omega$ invariant are $\sigma = 0, \pi$, which corresponds to the elements $1$ and $-1$ of the $U(1)_R$. Thus, $\Omega$ preserves a $\IZ_{2R}$ subgroup of the full $U(1)_R$. 
Consequently the fields $b^\Omega$ and $c^\Omega$ also preserve the $Z_{2R}$ subgroup of the full $U(1)_R$.
 An analogous analysis of the reduction ansatz of the metric leads to the fact that both $t$ and $\theta$ preserve a $\IZ_{2R}$ subgroup of the $U(1)_R$.}.  Therefore the 5d modes appearing in the entire truncation in table 2 is invariant under an $SU(2) \times SU(2) \times \IZ_2 \times \IZ_{2R}$. As we will see in the next section, the Klebanov-Strassler solution can be embedded in this truncation \cite{Cassani}. 

\begin{table}[h]
\centering
\begin{tabular}{c|c|c|c|c}
\hline 
 IIB fields & scalars & 1-forms & 2-forms & 5d metric \\ 
\hline 
10d metric & $u,v,t,\theta$ & $A$ &   & $g_{\mu\nu}$  \\		 

$\phi$ &  $\phi $ &   &   &   \\ 

B & $b^{\Phi},b^{\Omega}$ & & &  \\ 

$C_{0}$ & $C_{0}$ & & & \\

$C_{2}$ & $c^{\Phi},c^{\Omega}$ &  &  &  \\

$C_{4}$ & $a$ & $a_{1}^{J}$ &  & \\
\hline
\end{tabular}
\caption{Field content for the $ \mathcal{N}=2,$ $SU(2)\times SU(2)\times \mathbb{Z}_2$ truncation.}
\label{n=2trunc}
\end{table}

In the Klebanov-Strassler solution, the flux parameter $p$ and the following fields are consistently set to zero
\bea\label{nonKSfields}
\{ \real[b^{\Omega}] , \imag[c^{\Omega}] , a, C_{0} , c^{\Phi} , \theta ,  A, a_{1}^{J} \}~.
\eea
In this paper we will not consider perturbations of the Klebanov-Strassler solution by the above fields.  The fields that remain have the discrete $\IZ_2$ $R$-symmetry from before, and so are again part of an $SU(2)\times SU(2)\times\mathbb{Z}_2 \times \IZ_{2R}$ truncation\footnote{This can also be understood as a sub-truncation of the Papadopoulos-Tseytlin ansatz \cite{PT}.}.  For brevity, we will often refer to it as the $\IZ_2$-truncation as well: since this is the truncation we will work with exclusively in this paper, it should not cause any confusion with the full $\IZ_2$ truncation of the previous paragraph. We will study perturbations of the KS solution by scalar fields which are already activated in the background. The action governing these perturbations is given by \cite{Cassani}
\begin{align}\label{z2action}
\begin{split}
S_{b} &= \dfrac{1}{2\kappa_{5}^{2}}\int\biggl[R -\dfrac{28}{3} du^{2} -\dfrac{4}{3}dv^{2} -\dfrac{8}{3}du\,dv - dt^{2} - e^{-4u-\phi} \cosh 2t \, (db^{\Phi})^{2} \\
&\qquad  - \dfrac{1}{2} d\phi^{2}  + 2 \, e^{-4u-\phi} \sinh 2t \, db^{\Phi}d\boi -e^{-4u-\phi} \cosh 2t \, (db_{\mathcal{I}}^{\Omega})^{2}   \\
&\qquad- e^{-4u +\phi} (dc_{\mathcal{I}}^{\Omega})^{2}   - 4 e^{-\frac{20}{3}u + \frac{4}{3}v} + 24 \cosh t \, e^{-\frac{14}{3}u -\frac{2}{3}v}- 9 \sinh^{2}t \, e^{-\frac{8}{3}u -\frac{8}{3}v}  \\
&\qquad - 9 e^{-\frac{20}{3}u -\frac{8}{3}v -\phi} (b_{\mathcal{I}}^{\Omega})^{2}  -2e^{-\frac{32}{3}u -\frac{8}{3}v} \left( 3b_{\mathcal{I}}^{\Omega} c_{\mathcal{R}}^{\Omega} -q \, b^{\Phi} +k \right)^{2} \\
&\qquad - e^{-\frac{20}{3}u -\frac{8}{3}v +\phi} \, \left(9 ( c_{\mathcal{R}}^{\Omega})^{2}\cosh 2t - 6 q \; \cor \sinh 2t + q^{2} \cosh 2t \right) \Biggr]\star 1 ~,
\end{split}
\end{align}
where $\kappa_5$ is related to $\kappa_{10}$ as follows
\bea
\kappa_{5}^{2}= \dfrac{\kappa_{10}^{2}}{V_{Y}}~,\  ~~~~\text{where}~~~~~ V_{Y}= \dfrac{1}{2}\int_{T^{1,1}} J\wedge J\wedge \eta=\frac{16\pi^3}{27}~.
\eea
$V_{Y}$ is the unit volume of $T^{1,1}$. For later convenience we write down the metric on the scalar manifold in the basis
\begin{equation}\label{scalbasis}
\varphi^{I} = \{u,v,t,\phi,\bphi,\boi,\cor \}~,
\end{equation}
as follows
\bea\label{scalmetric}
\calg_{IJ} &=& \left(
\begin{array}{ccccccc}
 \frac{28}{3} & \frac{4}{3} & 0 & 0 & 0 & 0 & 0 \\
 \frac{4}{3} & \frac{4}{3} & 0 & 0 & 0 & 0 & 0 \\
 0 & 0 & 1 & 0 & 0 & 0 & 0 \\
 0 & 0 & 0 & \frac{1}{2} & 0 & 0 & 0 \\
 0 & 0 & 0 & 0 & e^{-4 u-\phi } \cosh 2 t & -e^{-4 u-\phi } \sinh 2 t & 0 \\
 0 & 0 & 0 & 0 & -e^{-4 u-\phi } \sinh 2 t & e^{-4 u-\phi } \cosh 2 t & 0 \\
 0 & 0 & 0 & 0 & 0 & 0 & e^{ -4 u +\phi} \\
\end{array}
\right)~.
\eea
With these definitions we can write the bosonic action as
\bea\label{z2bosonic}
S_{b} =\dfrac{1}{2\kappa_{5}^{2}} \int d^{5}x \sqrt{-g} \left(R -  \calg_{IJ} \partial_{A}\varphi^{I} \partial^{A}\varphi^{J} + \calv (\varphi) \right)~,
\eea
where $ A,B $ are indices for the spacetime coordinates and $ I,J$ index the scalar fields. The scalar potential $\cal V$, given by, 
\begin{align}\label{scalarpotential}
\begin{split}
{\cal V}(\varphi)&=- 4 e^{-\frac{20}{3}u + \frac{4}{3}v} + 24 \cosh t \, e^{-\frac{14}{3}u -\frac{2}{3}v}- 9 \sinh^{2}t \, e^{-\frac{8}{3}u -\frac{8}{3}v}  \\
&\qquad - 9 e^{-\frac{20}{3}u -\frac{8}{3}v -\phi} (b_{\mathcal{I}}^{\Omega})^{2}  -2e^{-\frac{32}{3}u -\frac{8}{3}v} \left( 3b_{\mathcal{I}}^{\Omega} c_{\mathcal{R}}^{\Omega} -q \, b^{\Phi} +k \right)^{2} \\
&\qquad - e^{-\frac{20}{3}u -\frac{8}{3}v +\phi} \, \left(9 ( c_{\mathcal{R}}^{\Omega})^{2}\cosh 2t - 6 q \; \cor \sinh 2t + q^{2} \cosh 2t \right) ~,
\end{split}
\end{align}
can be written in terms of a superpotential $\cal W$, given by
\bea\label{superpotential}
\mathcal{W} = e^{-\frac43  (4u+v)}\(3\,\boi \cor -q\,b^{\Phi} +k\)+3 \cosh t \, e^{-\frac43(u+v)}+2 e^{-\frac23(5u-v)}~,
\eea
as follows
\bea
\calv(\varphi) &=& -\calg^{IJ}\partial_{I}\mathcal{W} \partial_{J}\mathcal{W}+ \dfrac{4}{3}\mathcal{W}^{2}~.
\eea

Supersymmetric solutions of this system are obtained by analyzing the vanishing of the fermionic variations. The dimensional reduction of the 10d fermionic SUSY variations was performed in \cite{Liu:2011dw}. After converting their formulas into the notation of Cassani and Faedo (see appendix \ref{appd}) we obtain the fermionic variations listed in Appendix \ref{appd1}.

Among the 5d fields in Table \ref{n=2trunc}, the complex scalars $b^\Omega, c^\Omega$ and $z=\tanh t ~e^{i\theta}$ have R-charge 2 under the $U(1)$ R-symmetry of the boundary theory\footnote{A different way to see the R-charges is as follows: The holomorphic (2,0)-form $\Omega$ has a non-zero charge $q=-3$ under the action of the Reeb vector $\xi=3\partial_\psi$ (where the coordinate $\psi$ is defined in \eqref{oneforms1}). For a tensor $X$ its charge $q$ under the action of the Reeb vector is defined as ${\cal L}_\xi X=iqX $ (see for instance \cite{Ashmore:2016oug}). The R-charge $r$ is related to $q$ by $q=3r/2$. Therefore $\Omega$ has R-charge $-2$ which implies that $b^\Omega$ and $c^\Omega$ have R-charge $+2$. Alternatively, one can also look at the gauge covariant derivative of $b^\Omega, c^\Omega,z$ and read off $q=+3$.}. Setting these scalars to zero consistently gives rise to a further truncation to $SU(2)\times SU(2)\times\mathbb{Z}_2 \times U(1)_R$ invariant modes. For later convenience, we will refer to this sector as the $U(1)$ truncation. The resulting model was considered in \cite{Bertolini,Aharony,Buchel}. The model is comparatively simpler and the action reads
\begin{align}\label{u1action}
\begin{split}
S= \dfrac{1}{2\kappa_{5}^{2}}\int\biggl[R -&\dfrac{28}{3} du^{2} -\dfrac{4}{3}dv^{2} -\dfrac{8}{3}du\,dv - e^{-4u-\phi} \, (db^{\Phi})^{2} - \dfrac{1}{2} d\phi^{2} - 4 e^{-\frac{20}{3}u+ \frac{4}{3}v} \hspace{1cm} \\
& +24 \, e^{-\frac{14}{3}u -\frac{2}{3}v} -2e^{-\frac{32}{3}u -\frac{8}{3}v} \left(-q \, b^{\Phi} +k \right)^{2} - e^{-\frac{20}{3}u -\frac{8}{3}v +\phi} \, q^{2} \Biggr]\star 1~.
\end{split}
\end{align}
The action reduces to the form considered in \cite{Bertolini} with the following identification
\bea\label{UVcomb}
U = 4u + v~,~~~~ V = u - v~.
\eea
The fields $U$ and $V$ have the geometric interpretation of the breathing and squashing mode of $T^{1,1}$ respectively. We will, at times, use these linear combinations to compare with the notations of \cite{Bertolini}.

\section{Klebanov-Tseytlin vs Klebanov-Strassler: UV Asymptotics}\label{ktsection}

In this section, we present the Klebanov-Tseytlin (KT) and Klebanov-Strassler (KS) solutions in terms of the fields of the five-dimensional gauged supergravity theory discussed in the previous section. Both KS and KT are supersymmetric solutions and preserves $1/2$ of the ${\cal N}=2$ supersymmetry of the supergravity theory. We present the BPS equations and the explicit form of the solutions. We end this section with a comparison of the UV asymptotics of the two solutions. 
\subsection{Klebanov-Tseytlin solution}

The KT solution is a $1/2$ BPS solution and can be embedded in the $U(1)$ truncation \eqref{u1action}. From 5d point of view, the KT solution is a flat domain-wall where the 5d metric takes the following form
\bea\label{ktmetric}
&& ds_{5}^{2} = \dfrac{1}{z^{2}}\Biggl( e^{2X(z)} dz^{2} + e^{2Y(z)} \eta_{\mu\nu} dx^{\mu}dx^{\nu}\Biggr),
\eea
and the scalars are functions of the radial coordinate $z$ only. In the above parametrization of the metric the boundary is at $ z=0 $. The indices $ \mu,\nu$ run over $0,1,2,3$.  On this ansatz, the BPS equations resulting from the fermionic variations in Appendix \ref{appd3} take the following gradient flow form
\bea\label{bpsu1}
e^{-X(z)}z \partial_{z}\phi^{I}-\calg^{IJ}\partial_{J}\mathcal{W}=0~, \ \ \ e^{-X(z)}z \partial_{z} \log \left(\dfrac{e^{Y(z)}}{z}\right) + \dfrac{1}{3}\mathcal{W} = 0~.
\eea
The KT solution, which solves this set of BPS equations, is given by
\begin{align}\label{ktsol}
\begin{split}
t &= 0~,\  \ b^{\Omega}_{\mathcal{I}} \ = \ 0~, \ \ c^{\Omega}_{\mathcal{R}} \ =\  0~, \\ 
 \phi &= \log (g_{s})~, \ \ \ b^{\Phi} = -g_{s}q \, \log \biggl(\dfrac{z}{z_{0}}\biggr)~,\\ 
 X &= \dfrac{2}{3} \log(h_{\tx{KT}})~, \ \ Y=\dfrac{1}{6} \log(h_{\tx{KT}})~,\\ 
u &=  \dfrac{1}{4} \log(h_{\tx{KT}})~, \ \  v \ = \ \dfrac{1}{4} \log(h_{\tx{KT}})~,\\ 
h_{\tx{KT}}(z) &= \dfrac{1}{8} \biggl[ -4k + g_{s}q^{2} -4 g_{s} q^{2} \log \biggl(\dfrac{z}{z_{0}}\biggr) \biggr]~,
\end{split}
\end{align}
where  $ g_{s} $ is an integration constant, which, upon uplifting to 10d string theory becomes the string coupling constant. The independent flux parameters $k$ and $q$ are related to the number of regular and fractional branes respectively in the uplifted theory. $ z_{0} $ is an arbitrary scale introduced to make the argument of the log dimensionless. 

\subsection{Klebanov-Strassler solution}

The KS solution is a $1/2$ BPS solution of \eqref{z2action}. Unlike the KT solution, the KS solution cannot be embedded in the $U(1)$ truncation \eqref{u1action} because the $U(1)$ charged fields $t, \boi, \cor$ are activated in the KS solution. The KT solution in \eqref{ktsol} has a naked singularity at $ z_{s} $, such that $ h(z_{s}) =0 $, and and therefore cannot capture the full dynamics of the dual field theory. On the other hand, in the full ten-dimensional spacetime, the KS solution (which asymptotically matches the KT solution) is smooth in the IR \footnote{The five-dimensional compactification of the KS solution turns out to be singular in the IR. This can be deduced by calculating the curvature invariants and probing the IR limit. However this singularity (which is an artefact of dimensional reduction) is an acceptable singularity since it satisfies Gubser's criterion of good singularities \cite{Gubser:2000nd}.}. From 5d point of view the KS solution can be seen as a flat domain-wall where the metric takes the following form
\bea\label{ksmetric}
&& ds_{5}^{2} = e^{2\mathfrak{X}(\tau)} d\tau^{2} + e^{2\mathcal{Y}(\tau)} \eta_{\mu\nu} dx^{\mu}dx^{\nu}~. 
\eea
In this parametrization the boundary is at $\tau=\infty$. On this ansatz, the BPS equations take the following form
\bea\label{bpsu2}
e^{-\mathfrak{X}(\tau)} \partial_{\tau}\phi^{I}+\calg^{IJ}\partial_{J}\mathcal{W}=0, \ \ \ e^{-\mathfrak{X}(\tau)}\partial_{\tau}\mathcal{Y}(\tau) - \dfrac{1}{3}\mathcal{W} = 0.
\eea
The seemingly different relative sign compared to that of \eqref{bpsu1} is due to the fact that the boundary is at $ \tau=\infty $.

The KS solution, for this choice of metric is given by
\begin{align}\label{kssol}
& \cor = \frac{q \; \tau}{3 \sinh (\tau)}~,~ t = -\log \left(\tanh \left(\frac{\tau}{2}\right)\right)~,~ e^{2u}={3\over 2}h^{1/2}\epsilon^{4/3}K~\sinh\tau~,\\
&  e^{2v}={3\over 2}\frac{h^{1/2}\epsilon^{4/3}}{K^2}~,~ b^{\Phi} = \dfrac{g_{s}\; q\; \coth (\tau )}{3}  \big(\tau  \coth (\tau )-1\big)~,~\boi =\dfrac{g_{s} q \big(\tau \cosh (\tau)-\sinh (\tau)\big)}{3 \sinh ^2(\tau)}~,\non\\
& e^{2\mathfrak{X}}={1\over4}h^{4/3}\epsilon^{32/9}\({3\over2}\)^{2/3}K^{-4/3}\sinh^{4/3}\tau~,~~~e^{2\mathcal{Y}}=h^{1/3}\epsilon^{20/9}\({3\over2}\)^{5/3}K^{2/3}\sinh^{4/3}\tau~,\non
\end{align}
where
\bea
&& K(\tau)=\frac{\(\sinh(2\tau)-2\tau\)^{1/3}}{2^{1/3}\sinh\tau}~,~~~~h'(\tau)=-\alpha\frac{l(\tau)}{K^2(\tau)\sinh^2\tau}~,
\eea
and the function $l(\tau)$ is given by
\bea
l(\tau)=\frac{\tau \coth \tau-1}{4\sinh^2\tau}(\sinh 2\tau-2\tau)~.
\eea
The dilaton is constant in this solution and is given by $ \phi = \log (g_{s})$. In these formulas $ \epsilon $ is the conifold deformation parameter and $\alpha=(16 g_s q^2)/(81 \epsilon ^{8/3})$. The function $h(\tau)$ is the integral $h(\tau) = \int_{\infty}^{\tau}h'(x)\; dx$ which cannot be evaluated in a closed form.

To compare the KS solution with the KT solution, we first find the asymptotic relation between the radial coordinate $\tau$ in the KS metric \eqref{ksmetric} and the radial coordinate $z$ in the KT metric \eqref{ktmetric}. This relation is found to be \cite{Herzog:2002ih}
\bea\label{relcoord}
z^{2}= \dfrac{2^{5/3}}{3}\epsilon^{-4/3}  e^{-2\tau/3}~.
\eea
In the $z$ coordinate, the KS solution takes the following asymptotic form
\bea\label{asymptks}
t &=& 2a^{3}z^{3} + \calo(z^{9}),\  \ b^{\Omega}_{\mathcal{I}} \ = -\dfrac{2}{3}g_{s}q \(1+3\log(a\,z)\right) a^{3}z^{3}+ \calo(z^{9}) \ ,\\  
c^{\Omega}_{\mathcal{R}} &=& -2 q a^{3}z^{3}\log (a\,z) + \calo(z^{9}) , \ \ 
\phi \ = \ \log (g_{s}), \ \ \ b^{\Phi} = -\dfrac{g_{s}q}{3}-g_{s}q \, \log (a\, z) + \calo(z^{6}), \non\\ 
e^{2u}&=& h_{\tx{KS}}^{1/2}+\calo(z^{6})~, ~e^{2v}= h_{\tx{KS}}^{1/2}+\calo(z^{6})~,~ e^{2\mathfrak{X}} = \frac19 h_{\tx{KS}}^{4/3}+\calo(z^{6})~,~e^{2\mathcal{Y}}=\frac{1}{z^2} h_{\tx{KS}}^{1/3}+\calo(z^{4})~,\non
\eea
where
\bea\label{warpKS}
h_{\tx{KS}}(z)\ = \ -g_{s}q^{2}\biggl[\dfrac{1}{24}+\dfrac{1}{2}\log(a\, z) \biggr]+\calo(z^{6})~, \ \text{ and } \ a\ = \ \dfrac{3^{1/2}}{2^{5/6}} \epsilon^{2/3} .
\eea
The subleading terms are determined through the BPS equations \eqref{bpsu2} after replacing the $\tau$ derivatives with the $z$ derivatives using the asymptotic relation \eqref{relcoord}. The metric in \eqref{ksmetric}, under the coordinate change, is given by
\bea
ds_{5}^{2} &=& \dfrac{h_{\tx{KS}}^{4/3}}{z^{2}}\ dz^{2} +   \dfrac{h_{\tx{KS}}^{1/3}}{z^{2}} \ \eta_{\mu\nu}dx^{\mu} dx^{\nu} \equiv \dfrac{e^{2X}}{z^{2}} \ dz^{2} + \dfrac{e^{2Y}}{z^{2}} \ \eta_{\mu\nu}dx^{\mu} dx^{\nu},
\eea
where $ e^{2X} =h_{\tx{KS}}^{4/3} $ and $ e^{2Y} =h_{\tx{KS}}^{1/3} $. 

Plugging \eqref{relcoord} and the asymptotic expressions for $X$ and $Y$ found above in the KS metric \eqref{ksmetric}, we recover the form of the KT metric \eqref{ktmetric}. The comparison of $h_{\tx{KS}}$ with $h_{\tx{KT}}$ relates the flux parameter $k$ in terms of the flux parameter $q$, the conifold deformation parameter $\e$ and the scale $z_0$ introduced in \eqref{ktsol}. This relation, together with the $\mathbb{Z}_2$ symmetry, reflects the fact that the KS solution is dual to the symmetric point on the baryonic branch which exists only when $k$ is proportional to $q$. In contrast, the KT solution is more generic where $k$ and $q$ are independent parameters. On a related note, we furthermore see that although there is a smooth limit $(q\to 0)$ of the KT solution to the conformal Klebanov-Witten solution, there is no such limit for the the KS solution (since under $q\to 0$, $h_{\tx{KS}}\to0$). The baryonic branch of the deformed conifold has been discussed in \cite{GranaMinasian} and the mesonic branch in \cite{CKK2}.

\section{SUSY breaking perturbations of the KS solution}\label{sec:4}

In this section, we discuss the sub-leading perturbations of the KS solution by analyzing the full bosonic equations of motion and present the most general SUSY breaking deformation of KS upto order $z^4$. The equations of motion for the action \eqref{z2bosonic} are given by
\begin{align}\label{EOMs}
\begin{split}
& \dfrac{2}{\sqrt{-g}}\partial_{A} \left(\sqrt{-g} g^{AB} \calg_{IJ} \, \partial_{B} \varphi^{J}\right) + \dfrac{\partial \calv}{\partial \varphi^{I}} - \dfrac{\partial \calg_{JK}}{\partial \varphi^{I} } \partial_{A} \varphi^{J} \partial_{B} \varphi^{K} = 0~, \\
& R_{AB} = \calg_{IJ} \partial_{A} \varphi^{I} \partial_{B} \varphi^{J} - \dfrac{1}{3} g_{AB} \calv(\varphi)~.
\end{split}
\end{align}
We will take the flat domain-wall ansatz used in \eqref{ksmetric} for the metric which is supported by non-trivial profile for the seven scalars \eqref{scalbasis} along the radial direction. There are seven second order ordinary coupled differential equations coming from the scalar sector and two more from the $zz$ and $\m\n$ component of the Einstein's equation. We make the following ansatz for the asymptotic expansions\footnote{The particular parametrization for the scalars $U,V,X$ and $Y$ is motivated by a natural 10d uplift as explained in Appendix \ref{appa}.}
\begin{align}\label{UVExpAnsatz}
\begin{split}
&\varphi^{I}(z) = \varphi^{I}_{\tx{KS}} + \sum_{i=1}\left(C^{I}_{(i)} + D^{I}_{(i)} \log a z\right) z^{i}~,~~~~ \forall \ I \neq U,V ~, \\
& e^{U(z)} = h(z)^{\frac{5}{4}} \, h_{2}(z)\, h_{3}(z)^{4}~,~~~~  e^{V(z)} = h_{3}(z)\, h_{2}(z)^{-1}~,\\
& e^{X(z) } = h(z)^{\frac{2}{3}}\,h_{2}(z)^{\frac{1}{3}}h_{3}(z)^{\frac{4}{3}}~,~ \ e^{Y(z)} = h(z)^{\frac{1}{6}}\,h_{2}(z)^{\frac{1}{3}}h_{3}(z)^{\frac{4}{3}}~, 
\end{split}
\end{align}
where
\begin{align}\label{UVExpAnsatz1}
\begin{split}
&h(z) = h_{\tx{KS}} + \sum_{i=1}\left( h^{(1)}_{i} + h^{(2)}_{i} \log z \right)z^{i}~,\\
& h_{\a}(z) = 1 + \sum_{i=1}\left( C^{\a}_{(i)}  + D^{\a}_{(i)} \log az \right) z^{i}, \ \ \ \ \a=2,3~.
\end{split}
\end{align}
The subscript $\tx{KS}$ indicates the KS solution expanded around $ z=0 $ as given in \eqref{asymptks} and \eqref{warpKS}. 

Before presenting our asymptotic SUSY breaking solution, we make one technical comment about \eqref{UVExpAnsatz}. In setting up the power series ansatz, we have used a series expansion in $z$ (together with the logarithmic terms). If we were working with a conventional Fefferman-Graham gauge, only even powers of $z$ in the warp factors would be necessary. But since it is somewhat harder to capture the KS solution in the conventional Fefferman-Graham coordinate system, we prefer to keep both $X(z)$ and $Y(z)$ in the metric ansatz. For such a choice for the 5d metric, the series expansion with only even powers of $z$ was considered in \cite{Aharony}. We do not know of an argument why this is correct {\em a priori}, so we have kept the full expansion in all powers of $z$. If we work with terms that involve such odd powers of $z$, we will get solutions which are supported by coefficients appearing at linear order in the ansatz for the warp factors (one such coefficient is $ C^{(2)}_{(1)} $ appearing in \eqref{UVExpAnsatz1}). However, we find that such solutions are unphysical and can be gauged away by a redefinition of the radial coordinate\footnote{In Appendix \ref{appb}, we show this explicitly in pure AdS by showing that this mode can be gauged away by a redefinition of the radial coordinate.}. Apart from this gauge mode, all the terms that appear up to $\calo(z^4)$ are even powers (consistent with \cite{Aharony}) and have physical interpretations (either as parameters in the KS solution, or as SUSY-breaking parameters appearing in \cite{DKM, Bertolini, Kuperstein}).

Upon substituting the series expansions into the equations of motion and solving them order by order in the radial coordinate $z$, we find the solution presented in \eqref{KSSUSYb} below. 
\begin{subequations}\label{KSSUSYb}
\begin{align}
\phi &= \log g_{s}+ \left(\varphi + 3\cals \log az\right)a^4z^{4} + \calo(z^{6})~,\\
\bphi &= -\dfrac{1}{3}g_{s}q-g_{s}q \log az + \frac{g_{s}q}{16} \Big(7\cals -4\varphi -24 \cals\log az\Big)  a^4z^{4} +\calo(z^{6})~,\\
\boi &= -\left(\dfrac{2}{3}g_{s}q + 2 g_{s}q \log a z \right)  a^{3}z^{3} + \calo(z^{6})~,\\
\cor &=  -2 q \log a z \, a^{3}z^{3} + \calo(z^{6})~,\\
t &= 2a^{3}z^{3} + \calo(z^{6})~,\\
h &= -\dfrac{\gs q^{2}}{24} (1+12 \log az) + \dfrac{\gs q^{2}}{192}\Big(35 \cals - 12\varphi -48 \cals \log az \Big)a^4z^{4} + \calo(z^{6})~,\\
h_{2} &= 1+ \dfrac{1}{2}\cals \, a^4z^{4}+ \calo(z^{6})~,\\
h_{3} &= 1+ \calo(z^{6})~.
\end{align}
\end{subequations}
Up to order $ z^{4} $ and $z^4 \log z$, the solution is determined by two independent, SUSY-breaking, integration constants ${\cal S}$ and $\varphi$. There are no new SUSY-breaking integration constants with respect to SUSY-breaking deformations of the KT solution studied in \cite{DKM,Bertolini}. The authors of \cite{BGGHM} found the most general deformation of the KS solution by considering the $SU(2)\times SU(2)\times \Z_2$ invariant Papadopoulos-Tseytlin ansatz in the Type IIB supergravity. Our finding is consistent with their result in that the subleading perturbations are characterized by a two parameter family of SUSY-breaking integration constants. We also find a number of SUSY-preserving integration constants. However we have set them to zero as they do not play any role in subsequent sections\footnote{Some of the additional integration constants are related to reparametrization of the radial coordinate (see Appendix \ref{appb} for illustration in pure AdS). Therefore discussion about its SUSY might seem unnecessarily pedantic. But the principle of setting SUSY-preserving parameters to zero is a more generally a useful idea. In Appendix \ref{appa}, we discuss the details of a more general ansatz and count the number of SUSY-preserving/breaking parameters in them.}. 

\section{Holographic Ward identities}\label{sec:5}

We would like to associate the SUSY breaking solution found in the previous section with a SUSY breaking vacua of the Klebanov-Strassler gauge theory. Since the KS theory is an ${\cal N}=1$ supersymmetric gauge theory, supersymmetry Ward identities should hold in any of its vacua. In this section we will derive the SUSY ward identities holographically and check them against the solution found in \eqref{KSSUSYb}. We we also derive other operator identities involving the trace of the energy-momentum tensor and $\gamma$-trace of the supercurrent. As we will see, these identities can be derived from relations between one-point functions of operators at generic sources \cite{Bertolini}. Therefore, we begin by identifying the holographic sources for dual operator and defining the one-point functions.

\subsection{Sources and dual operators}

In order to find the sources for the operators of the dual gauge theory, we study the equations of motion linearized around the asymptotic KS solution \eqref{asymptks}. In the superconformal Klebanov-Witten theory, the usual AdS/CFT correspondence dictates that fields of a certain mass $m$ in the bulk are dual to gauge invariant operators in the CFT of a certain conformal dimension $\Delta$. The mass-dimension relation depends upon the spin of the fields/operators. In Table \ref{massSpectrum} below we present the mass/dimension of fields/operators that are present in the $SU(2)\times SU(2) \times \Z_2$ truncation. 

\begin{table}[h]
\begin{center}
$	\begin{array}{lcccccccl}\hline
\mathcal N=2\: {\rm multiplet} && {\rm field\: fluctuations} && \hskip-0.3cm\text{AdS mass}   &&  {\rm spin} && \Delta \\ \hline
\rule{0pt}{3ex}
{\rm gravity} && \begin{array}{c}(A+2a_1^J)_A\\ \Psi_{A} \\ g_{AB} \end{array} && \hskip-0.3cm\begin{array}{l} m^2=0 \\m={3\over 2} \\ m^2=0 \end{array}  &&  \begin{array}{l} 1 \\ {3\over 2} \\ 2 \end{array} && \begin{array}{l} 3 \\{7\over 2} \\ 4 \end{array}  \\ \hline
{\rm universal \;hyper} && \begin{array}{c} b^\Omega + i \,c^\Omega \\ \z_\f\\ \tau=C_0+ie^{-\f} \end{array} && \begin{array}{l} m^2=-3 \\m=-{3\over 2}\\ m^2=0 \end{array} &&  \begin{array}{l} 0 \\ {1\over 2} \\ 1 \end{array} && \begin{array}{l} 3 \\{7\over 2} \\ 4 \end{array}  \\ \hline
{\rm Betti \;hyper} &&	\begin{array}{c}	 t \,e^{i\theta}\\ \z_b\\ b^\F,\;\;c^\F \end{array} && \begin{array}{l}m^2= -3 \\m=  -{3\over 2}\\m^2= 0   \end{array} &&  \begin{array}{l} 0 \\ {1\over 2} \\ 1 \end{array} && \begin{array}{l} 3 \\{7\over 2} \\ 4 \end{array}  \\ \hline
{\rm massive \;vector} && \begin{array}{c} V\\ \z_V\\(A-a_1^J)_A\\ b^\Omega - i \,c^\Omega \\ \z_U\\ U \end{array} && \begin{array}{l} m^2=12\\m={9\over2} \\m^2= 24 \\m^2= 21\\m=-{11\over2}\\m^2= 32  \end{array}   &&  \begin{array}{l} 0 \\ {1\over 2} \\ 1 \\0\\\frac12\\0\end{array} && \begin{array}{l} 6 \\{13\over 2} \\ 7\\7\\{15\over 2}\\8 \end{array}  \\ \hline
\end{array}$
\caption{Mass spectrum of bosons and fermions in the ${\cal N}=2,~\Z_2$ truncation of \cite{Cassani} around the supersymmetric $AdS_5$. In our conventions, setting $k= -2$ leads to a unit AdS radius (5d indices  are dubbed $A,B$).}
\label{massSpectrum}
\end{center}
\end{table}
All fields in this table are organized in multiplets of $5d$, ${\cal N}=2$ supersymmetry.  All the fermions are in the Dirac representation. The gravity multiplet contains the metric, a $U(1)$ vector field and the gravitino which comprises 8 bosonic and 8 fermionic on-shell real degrees of freedom. The hypermultiplets contains four real scalars and a Dirac fermion which comparises of 4 bosonic and 4 fermionic on-shell real degrees of freedom. The massive vector multiplet can be though as a massless vector that has undergone a Higgs mechanism by eating up an entire hypermultiplet. It contains 8 bosonic and 8 fermionic on-shell real degrees of freedom. To sum up, the matter content of the $\Z_2$ truncation can be seen as consisting of one vector multiplet and three hypermultiplets (splitting the massive vector into a massless vector and a hypermultiplet is convenient for writing down supersymmetry transformation rules). 
\subsubsection{Bosonic sector}
For bulk scalar fields which lie outside the double quantization window (as it is the case here), the non-normalizable mode is interpreted as a source for the dual operator. Since the bosonic scalar operators in the Table \ref{massSpectrum} have integer scaling dimensions, the linearized equations of motion for these fields around pure $AdS_5$ is solved by integer power law solutions. When we move to the KT/KS background, these power law solutions will get corrected by logarithmic terms (which capture the log running of the gauge coupling in the dual theory). With this in mind we start with an ansatz dictated by the pure AdS solution and add to it logarithmic terms. For convenience in the linearization procedure, we introduce a book-keeping parameter $ \varepsilon $ in the following ($n$ is not summed over in the following formulas)
\begin{align}
\begin{split}
&\varphi^{I}(z) = \varphi^{I}_{\tx{KS}} +\varepsilon\left(\delta\varphi^{I}_{(n)} + \delta\tilde{\varphi}^{I}_{(n)} \log a z\right) z^{n}~, ~~~~ \forall \ I \neq U,V~, \,\\
& e^{X(z) } = h(z)^{\frac{2}{3}}\,h_{2}(z)^{\frac{1}{3}}h_{3}(z)^{\frac{4}{3}}~, ~~ e^{Y(z)} = h(z)^{\frac{1}{6}}\,h_{2}(z)^{\frac{1}{3}}h_{3}(z)^{\frac{4}{3}}~, \\
& e^{U(z)} = h(z)^{\frac{5}{4}} \, h_{2}(z)\, h_{3}(z)^{4}~,  ~~~ \ e^{V(z)} = h_{3}(z)\, h_{2}(z)^{-1}~,
\end{split}
\end{align}
where
\begin{align}
\begin{split}
& h(z) = h_{\tx{KS}}^{0} + \varepsilon\left( \delta h_{(n)} + \delta \tilde{h}_{(n)} \log az \right)z^{n}~,\\
& h_{a}(z) = 1 + \varepsilon \left(\delta h^{(a)}_{(n)}  + \delta \tilde{h}^{(a)}_{(n)} \log az \right) z^{n}~, \ \ \ \ a=2,3~.
\end{split}
\end{align}
We will be interested in the following values of $n$:  $-4,-3,-2,0,1$. $n=1$ corresponds to the most relevant scalar operator of dimension three and $n=-4$ corresponds to the most irrelevant scalar operator of dimension eight in the theory. We solve the system separately for each $n$. In the presence of irrelevant operators, finding a solution to the full non-linear equations involving all the sources is an ill-defined problem \cite{Aharony}.

 The solution presented below corresponds to sources for all scalar operators of a given dimension turned on one at a time.
\begin{subequations}\label{holsources}
\bea
(i)&& \delta h =\delta h_{(-4)}(x)z^{-4}~,\\
(ii)&& \delta\boi =\delta\boi {}_{(-3)}(x)z^{-3}~, \ \ \delta\cor = -g_s^{-1}\delta\boi {}_{(-3)}(x)z^{-3}~, \\
(iii) && \delta h = \frac12\gs q^{2}\; \delta h^{(2)}_{(-2)}(x)z^{-2}~, \ \ \delta h_{2} = \delta h^{(2)}_{(-2)}(x)z^{-2}~, \ \ \delta h_{3} = -\frac14 \delta h^{(2)}_{(-2)}(x)z^{-2}~,\non\\
&& \delta\bphi = -\frac12\gs q \; \delta h^{(2)}_{(-2)}(x)z^{-2}~, \\
(iv)&&  \delta\bphi =  \delta\bphi_{(0)}(x) -g_{s}q  \delta\phi_{(0)}(x) \log az~,  \ \ \delta\phi =  \delta\phi_{(0)}(x)~, \ \\
&& \delta h = \dfrac{1}{8} \(4 q \delta\bphi_{(0)}(x)+g_{s}q^{2} \delta\phi_{(0)}(x) -4 g_{s}q^{2}\delta\phi_{(0)}(x) \log az \)~,\non \\
(v) && \delta t= \dfrac{1}{\gs q}\(\delta b_{(1)}+\delta t_{(1)} \log a z\)z~,~~\delta \cor = \frac{ 1}{24\gs} \[18 \delta b_{(1)} + \delta t_{(1)} \(1+12 \log az\)\]z~,\non\\
&&\delta \boi = \frac{1}{12} \( 3 \delta b_{(1)} + 2\delta t_{(1)} \)z~.
\eea
\end{subequations}

In the solutions listed above, the first three correspond to sources for irrelevant operators of dimensions eight, seven and six respectively. The solution in $ (iv) $ contains the source for the two marginal scalar operators (that corresponds to the sum and the difference of the gauge couplings). The solutions in $ (v) $ contain sources for operators of dimension three which are new in the $\Z_2$ truncation. These sources corresponds to gaugino mass terms and therefore break supersymmetry explicitly.

In the metric sector, we have the transverse-traceless fluctuations of the metric induced on a finite radial cut-off surface, which in the boundary limit, sources the energy-momentum tensor of the boundary theory. The induced metric at a finite radial cut-off is given by
\bea
\gamma_{\mu\nu} &=& e^{2Y}\ol{\gamma}_{\mu\nu}, \ \ \text{where} \ \ \ol{\gamma}_{\mu\nu} = \dfrac{\eta_{\mu\nu}}{z^{2}}~.
\eea
The independent source from the metric which decouples from the rest of the sources is then given by
\bea\label{metricsource}
\delta \ol{\gamma}_{\mu\nu } = \dfrac{\delta h_{\mu\nu}(x)}{z^{2}}~.
\eea

Having obtained the sources, we now give the field operator map. The $SU(2)\times SU(2)\times \Z_2$ invariant sector of gauge invariant operators in the Klebanov-Strassler theory are, in general, dual to bulk fields which are composite. The two marginal operators ${\cal O}_+\equiv \Tr\(F_{(1)}^2+F_{(2)}^2\)$ and ${\cal O}_-\equiv\Tr\(F_{(1)}^2-F_{(2)}^2\)$ (that correspond to the sum and the difference of the gauge coupling) are dual to $e^{-\phi}$ and $e^{-\phi}b^\Phi$ respectively \cite{KS} whereas the two relevant operators ${\cal Q}_+\equiv \Tr\(W_{(1)}^2+W_{(2)}^2\)$ and ${\cal Q}_-\equiv \Tr\(W_{(1)}^2-W_{(2)}^2\)$ (that correspond to mass terms of the gaugino bilinears) are dual to the combination $b^\Omega+ig_s c^\Omega$ and $t$ respectively \cite{Loewy:2001pq,Kuperstein:2003yt}\footnote{In their convention, the linear combination dual to $ {\cal Q}_+ $ is $ b^\Omega - ig_s c^\Omega $.}. The sources in \eqref{holsources} and \eqref{metricsource} corresponds to these operators as follows
\begin{equation}\label{sourceOp}
{\cal O}_+ \leftrightarrow \delta\phi_{(0)}~,~~~~{\cal O}_- \leftrightarrow \delta\bphi_{(0)}~,~~~~{\cal Q}_+ \leftrightarrow \delta b_{(1)}~,~~~~{\cal Q}_- \leftrightarrow \delta t_{(1)}~,~~~~T_{\m\n} \leftrightarrow \delta h_{\mu\nu}~.
\end{equation}

The sources obtained in \eqref{holsources} are not diagonal by which we mean that a mode for one field can simultaneously turn on multiple fields. For example a non-zero $ \delta\phi_{(0)} $ results in turning on $ \delta\phi $ and $ \delta\bphi $. On the other hand, the composite field $e^{-\phi}b^\Phi$ is not affected by $ \delta\phi_{(0)} $ (it is turned on by $\delta\bphi_{(0)}$ only). This, however, is not true for the sources of dimension three operators. Regardless, we find it convenient to define combinations which are diagonal in the sources as it will be relevant later when we consider supersymmetry transformation of the sources.
\begin{align}\label{diagSources}
\begin{split}
&\delta \ol{\phi} = \delta \phi~,~~~~~~ \delta \ol{b}^\Phi = \delta\bphi + \gs q~\delta \phi \log a z ~,\quad \quad\delta \ol{t} =  \dfrac{24}{5-12 \log az}\Big(\delta B_+ -\gs q \delta t\Big) ~,\\
&\delta \ol{B}_{+} = \dfrac{1}{5-12 \log az}\Big( -24 \log az \delta B_+ +\gs q (5+12 \log az)\delta t\Big) ~,
\end{split}
\end{align}
where we have defined $ \delta B_{+} = \delta\boi+ \gs \delta\cor $. All the hatted  fields are sourced independently\footnote{In \cite{Bertolini}, where fields dual to marginal operators only mattered, analogous relations were written down for the composite fields by re-writing the explicit $z$-dependencies on the right hand sides in terms of bulk fields.} i.e., 
\begin{align}\label{diagSources1}
\begin{split}
 \delta\ol{\phi}  &= \delta \phi_{(0)}~,~~~~~~ \delta\ol{b}^\Phi  = \delta \bphi_{(0)}~,\\
 \delta \ol{B}_{+} &=  ~ \delta b_{(1)} z ~,~~~~ \delta \ol{t} = ~ \delta t_{(1)} z  ~.
 \end{split}
\end{align}
The holographically renormalized one point functions of the marginal operators in \eqref{sourceOp} was first obtained in \cite{Aharony} by functionally differentiating the on-shell renormalized action w.r.t. the corresponding sources in \eqref{sourceOp}. The renormalized one-point functions are then obtained by taking appropriate boundary limits. In the following we give an independent derivation of these one-point functions (including the dimension three operators which are new) by taking a slightly different approach where, the renormalized one point functions are obtained by functionally differentiating the on-shell renormalized action w.r.t. to the hatted (composite induced) fields and taking appropriate limits. The two procedure are equivalent in AAdS background but as we will see later in the derivation of the Ward identities (section \ref{sec5.2}), the latter definition is crucial in the KS background. We have 
\begin{align}\label{oneptfn1}
\begin{split}
\lag T_{\m\n}\rag &=\frac{2}{\sqrt{-\hat{\gamma}}} \frac{\delta S_{\tx{ren}}}{\delta h^{\m\n} }= \frac{2}{\sqrt{-\hat{\gamma}}} \frac{\delta S_{\tx{ren}}}{\delta \gamma_{\rho\sigma} }\frac{\delta \gamma_{\rho\sigma}}{\delta h^{\m\n} }~,\\
\lag {\cal O}_+\rag &= \frac{1}{2\sqrt{-\hat{\gamma}}}\frac{\delta S_{\tx{ren}}}{\delta\hat{\phi} } =\frac{1}{2\sqrt{-\hat{\gamma}}}\frac{\delta S_{\tx{ren}}}{\delta\phi_{(0)} } \\
&= \frac{1}{2\sqrt{-\hat{\gamma}}}\bigg[\frac{\delta S_{\tx{ren}}}{\delta\phi }\frac{\delta \phi}{\delta\phi_{(0)} }+\frac{\delta S_{\tx{ren}}}{\delta\bphi }\frac{\delta \bphi}{\delta\phi_{(0)} }+\frac{\delta S_{\tx{ren}}}{\delta U }\frac{\delta U}{\delta\phi_{(0)} }+\frac{\delta S_{\tx{ren}}}{\delta \gamma_{\m\n} }\frac{\delta \gamma_{\m\n}}{\delta\phi_{(0)} }\bigg]~,\\
\lag {\cal O}_-\rag &= \frac{1}{2\sqrt{-\hat{\gamma}}}\frac{\delta S_{\tx{ren}}}{\delta\hat{b}^\Phi}  =\frac{1}{2\sqrt{-\hat{\gamma}}}\frac{\delta S_{\tx{ren}}}{\delta\bphi_{(0)} }=  \frac{1}{2\sqrt{-\hat{\gamma}}}\bigg[\frac{\delta S_{\tx{ren}}}{\delta\bphi }\frac{\delta \bphi}{\delta\bphi_{(0)} }+\frac{\delta S_{\tx{ren}}}{\delta U }\frac{\delta U}{\delta\bphi_{(0)} }+\frac{\delta S_{\tx{ren}}}{\delta \gamma_{\m\n} }\frac{\delta \gamma_{\m\n}}{\delta\bphi_{(0)} }\bigg]~,\\
\lag {\cal Q}_+\rag &=\frac{1}{2\sqrt{-\hat{\gamma}}}\frac{\delta S_{\tx{ren}}}{\delta\hat{B}_{+} } =\frac{1}{2\sqrt{-\hat{\gamma}}}\frac{\delta S_{\tx{ren}}}{\delta b_{(1)}z }= \frac{1}{2\sqrt{-\hat{\gamma}}}\bigg[\frac{\delta S_{\tx{ren}}}{\delta t }\frac{\delta t}{\delta b_{(1)} }+\frac{\delta S_{\tx{ren}}}{\delta \boi }\frac{\delta \boi}{\delta b_{(1)} }+\frac{\delta S_{\tx{ren}}}{\delta \cor }\frac{\delta \cor}{\delta b_{(1)} }\bigg]~,\\
\lag {\cal Q}_-\rag &=\frac{1}{2\sqrt{-\hat{\gamma}}}\frac{\delta S_{\tx{ren}}}{\delta\hat{t} } =\frac{1}{2\sqrt{-\hat{\gamma}}}\frac{\delta S_{\tx{ren}}}{\delta t_{(1)} z }= \frac{1}{2\sqrt{-\hat{\gamma}}}\bigg[\frac{\delta S_{\tx{ren}}}{\delta t }\frac{\delta t}{\delta t_{(1)} }+\frac{\delta S_{\tx{ren}}}{\delta \boi }\frac{\delta \boi}{\delta t_{(1)} }+\frac{\delta S_{\tx{ren}}}{\delta \cor }\frac{\delta \cor}{\delta t_{(1)} }\bigg]~.
\end{split}
\end{align}
In these formulas, $S_{\tx{ren}}$ is the renormalized on-shell action given by $S_{\tx{ren}}=S_{\tx{reg}}+S_{\tx{ct}}$, where $S_{\tx{reg}}$ is the regulated action computed at a finite radial cut-off and $S_{\tx{ct}}$ is the counterterm action. Using \eqref{holsources}, these expressions can be simplified to the following
\begin{align}\label{oneptfn2}
\begin{split}
\lag T_{\mu\nu} \rag &= \dfrac{2}{\sqrt{-\ol{\gamma}}} \,  \dfrac{\de S_{\tx{ren}}}{\de \gamma^{\mu\nu}}\hks^{1/3}~,\\
\lag {\cal O}_+\rag &= \dfrac{1}{2\sqrt{-\ol{\gamma}}} \bigg[ \dfrac{\de S_{\tx{ren}}}{\de \phi} - \gs q \la \dfrac{\de S_{\tx{ren}}}{\de \bphi} + \Big( 1+ \dfrac{\gs q^{2}}{6\hks}\Big)\bigg( \dfrac{5}{4} \dfrac{\de S_{\tx{ren}}}{\de U} + \dfrac{1}{3} \gamma_{\mu\nu}\dfrac{\de S_{\tx{ren}}}{\de \gamma_{\mu\nu}} \bigg)\bigg]~,\\
\lag {\cal O}_-\rag &= \dfrac{1}{2\sqrt{-\ol{\gamma}}} \bigg[ \dfrac{\de S_{\tx{ren}}}{\de \bphi} + \dfrac{q}{2\hks}\bigg( \dfrac{5}{4} \dfrac{\de S_{\tx{ren}}}{\de U} + \dfrac{1}{3} \gamma_{\mu\nu}\dfrac{\de S_{\tx{ren}}}{\de \gamma_{\mu\nu}} \bigg) \bigg]~,\\
\lag {\cal Q}_+\rag &= \dfrac{1}{2\sqrt{-\ol{\gamma}}} \bigg[\frac{1}{4} \dfrac{\de S_{\tx{ren}}}{\de \boi} +\frac{3}{4\gs}  \dfrac{\de S_{\tx{ren}}}{\de \cor}  + \frac{1}{\gs q} \dfrac{\de S_{\tx{ren}} }{\de t} \bigg]~,\\
\lag {\cal Q}_-\rag &= \dfrac{1}{2\sqrt{-\ol{\gamma}}} \bigg[\frac{1}{6} \dfrac{\de S_{\tx{ren}}}{\de \boi} +\frac{1}{24\gs}\(1+12 \la\)  \dfrac{\de S_{\tx{ren}}}{\de \cor}  +\frac{1}{\gs q } \la \dfrac{\de S_{\tx{ren}} }{\de t} \bigg]~.
\end{split}
\end{align}
The renormalized QFT one-point functions of these operators are obtained by taking the following limits
\begin{align}\label{oneptfn3}
\begin{split}
\lag T^\m_{~\n}\rag_{\tx{QFT}}= &\lim\limits_{z\rightarrow 0 } z^{-4}~\lag T^\m_{~\n}\rag~,~~~~\lag {\cal O}_+\rag_{\tx{QFT}}= \lim\limits_{z\rightarrow 0 } z^{-4}~ \lag {\cal O}_+\rag~,~~~~\lag {\cal O}_-\rag_{\tx{QFT}}= \lim\limits_{z\rightarrow 0 } z^{-4}~ \lag {\cal O}_-\rag~,\\
&~~~~\lag {\cal Q}_+\rag_{\tx{QFT}}= \lim\limits_{z\rightarrow 0 } z^{-3}~ \lag {\cal Q}_+\rag~,~~~~\lag {\cal Q}_-\rag_{\tx{QFT}}= \lim\limits_{z\rightarrow 0 }z^{-3}~ \lag {\cal Q}_-\rag~.
\end{split}
\end{align}

\subsubsection{Fermionic sector}

The fermionic content of the full $ SU(2) \times SU(2) \times \Z_2$ truncation of $ \caln=2 $ supergravity is made up of a gravitino $\Psi_\m$, three hyperinos $\zeta^A$ ($A=1,2,3$) and a gaugino\footnote{$u_3$ is the scalar that appears in the `massless' vector multiplet. See discussion above \eqref{vsmMetric} and Eq. \eqref{bertolinimap} for a clarification on the notation used here.} $\lambda^{u_3}$. A detailed discussion of the fermionic sector and the supersymmetry of the $SU(2) \times SU(2) \times \Z_2$ truncation is given in Appendix \ref{appd}, including the mapping of notations used in \cite{Bertolini, Halmagyi:2011yd,Liu:2011dw}.

The equations of motion for the fermions and the gravitino were originally obtained in \cite{Liu:2011dw}. To obtain the sources of the dual fermionic operators, we first project the fermions onto definite chirality (which is well-defined at a given radial surface) and then solve the equations in the asymptotic KS background given in \eqref{asymptks}. We can make a crucial observation at this stage, by looking at the equations of motion in \cite{Liu:2011dw}: if we repackage the bosonic background in the equations of motion by powers of $ z $, the leading terms are sensitive only to the $ \calo(1) $ and $ \calo(\log az) $ terms of the bosonic fields. What this means is that the leading order terms in the first order differential equations are identical to that one finds in the KT background, with an appropriate identification of the parameters $ k $ and $ a $. The solutions to the equations of motion in the KT background have been found by \cite{Bertolini},
\begin{subequations}
\bea
\zeta_{\phi}^{-} &=& \sqrt{z}\, \hks(z)^{-\frac{1}{12}}\, \psi_{1}^{-}(x) + \calo(z^{\frac{3}{2}}),\\
\zeta_{b}^{-} &=& \dfrac{\sqrt{z} \hks(z)^{-\frac{1}{12}}}{20 q} (24\hks(z) - 5g_{s}q^{2}) \psi_{1}^{-}(x) +  \sqrt{z} \hks(z)^{-\frac{3}{4}}\psi_{2}^{-}(x) + \calo(z^{\frac{3}{2}}),\\
\zeta_{U}^{-} &=& \dfrac{3}{4}\sqrt{z}\hks(z)^{-\frac{1}{12}}\psi_{1}^{-}(x) +\dfrac{5q}{8} \sqrt{z} \hks(z)^{-\frac{7}{4}} \psi_{2}^{-}(x) + \calo(z^{\frac{3}{2}}), \\
\zeta_{V}^{+} &=&  \calo(z^{\frac{3}{2}}) ,\\
\Psi_{\mu}^{+} &=& \dfrac{\hks(z)^{\frac{1}{12}}}{\sqrt{z}} \psi^{+}_{\mu}(x) + i \dfrac{\hks(z)^{\frac{1}{6}}}{g_{s}q^{2}\sqrt{z}}\gamma_{\mu}  \bigg(-\dfrac{4}{5}\hks(z)^{\frac{11}{12}}\psi_{1}^{-}(x)  \non\\
&&\qquad\qquad\qquad\qquad\qquad + q \hks(z)^{-\frac{7}{4}}\(\hks(z)+\dfrac{g_{s}q^{2}}{12}\)\psi_{2}^{-}(x)\bigg)+ \calo(z^{\frac{3}{2}}).\qquad\qquad
\eea
\end{subequations}
For completeness, we restate some crucial comments regarding these solutions below. In pure AdS $ \zeta_{\phi} $ and $ \zeta_{b} $ have masses $ m_{\phi,b}=-3/2 $, $ \zeta_{U} $ has mass $ m_{U} = -15/2 $, $ \zeta_{V} $ has mass $ m_{V} =11/2 $ and the gravitino has a mass $ m_\Psi =3/2 $. One basic idea behind solving the equations of motion for Dirac fields in $AdS_5$ is that a Dirac spinor in five dimensions has the same number of components as a Dirac spinor in four dimensions. However a Dirac spinor in 5d is irreducible while in 4d it is reducible and the minimal spinors are Weyl spinors which contain half as many physical degrees of freedom. Since the boundary operators are of definite chirality, it is imperative to decompose the 4d projection of the 5d spinors onto a definite chirality. The two chiralities at a given radial slice have different UV fall-offs and is determined by the sign of the fermion mass term (see \cite{Henneaux} for a detailed discussion). Following this, the leading chirality of $ \zeta_{\phi},\zeta_{b}$ are negative, while for $ \Psi_{\mu} $ is positive. We don't consider the sources for the irrelevant operators dual to $ \zeta_{U}^{-} $ and $ \zeta_{V}^{+} $.

Now, we will focus our attention on finding the fermionic superpartners of the composite bosonic fields. We will only be focussing on the sources $ \hat{\phi} $, $ \hat{b}^\Phi $ the metric fluctuation $ h_{\mu\nu} $ and their fermionic superpartners, as these are the only inputs required in the SUSY Ward identity computation\footnote{This point is further elaborated with reasons in the subsection where we compute the Ward identities.}.  To this effort, we can use the following relation
\bea
\delta_{\epsilon}\hat{\varphi} = \dfrac{\de \hat{\varphi}}{\de \varphi^{I}} \delta_{\epsilon} \varphi^{I},
\eea
and use the SUSY variations of the bosonic fields given in \eqref{bosonSUSYvar}. These relations will be useful in computing the SUSY Ward identities. We also need the following relations, where this relation is evaluated in the KS background, 
\bea
\delta_{\epsilon}\hat{\phi} &=& \dep\phi, \\
\dep \hat{b}^\Phi &=& \dep\bphi+ \gs q \log az \, \dep\phi,\\
\dep \hat{e}^{a}_{\mu} &=& \hks^{-1/6}\bigg(\dep e^{a}_{\mu} - \dfrac{1}{48\hks}e^{a}_{\mu}(4q \dep \bphi + \gs q^{2}\dep \phi)  \bigg)
\eea

Using the KS background and the bosonic SUSY variations in \eqref{bosonSUSYvar}, along with \eqref{bertolinimap}, we can write
\begin{subequations}
\bea\label{compferm}
\dep \hat{\phi} &\equiv& \dfrac{i}{2} \big( \bar{\epsilon}\hzd{\phi} -\hzdb{\phi} \epsilon \big) \ = \ \dfrac{i}{2}\big(\bar{\epsilon} \zeta_{\phi} - \bar{\zeta}_{\phi} \, \epsilon \big),\\
\dep \hat{b}^\Phi &\equiv& \dfrac{i}{2} \big( \bar{\epsilon}\hzd{b} -\hzdb{b} \epsilon \big) \ = \ \dfrac{i}{2} \big( \bar{\epsilon} \zeta_{b} - \bar{\zeta}_{b} \, \epsilon  \big) +  \dfrac{i}{2}\gs q \log az\big(\bar{\epsilon} \zeta_{\phi} - \bar{\zeta}_{\phi} \, \epsilon \big) +  \dots,\\
\label{compgraviton}\dep \hat{e}^{a}_{\mu} &\equiv& \dfrac{1}{2} \bar{\epsilon} \gamma^{a}\ol{\Psi}_{\mu} +\text{ h.c.} \ = \ \dfrac{\hks^{-1/6}}{2}\bar{\epsilon}\gamma^{a}\Psi_{\mu}  + \text{h.c.} + \dots,
\eea
\end{subequations}
where the dots indicate subleading terms, which are suppressed by powers of $ z $ or by factors of $ \log az $ in the denominator. In the rest of the paper, we will use the hatted fermions to indicate the combinations, to leading order, of the original supergravity fermions that are defined above. The subleading terms do not contribute when we take the $ z\rightarrow 0 $ limit.

Although we do not require the explicit form of the fermionic action for the purpose of computing the SUSY Ward Identities, we need the formal prescription for computing the one point functions, and taking the boundary limit. The one point functions of the fermions dual to the composite bosons and their boundary limits are defined as follows
\bea\label{foneptfundef}
\begin{split}
\lr{\bar{S}^{\nu -}} &=& \dfrac{-2i}{\sqrt{-\ol{\gamma}}} \dfrac{\delta S_{f,\tx{ren}}}{\delta \hat{\Psi}^{+}_{\nu}} , \qquad \lr{\bar{S}^{\nu -}}\qft = \lim\limits_{z\rightarrow0} z^{-\frac{9}{2}}\hks^{-\frac{1}{12}}\, \lr{\bar{S}^{\nu -}}\\
\lr{\bar{\mathcal{O}}^{+}_{\hat{\zeta}_{\ol{\phi}}}} &=& \dfrac{i}{\sqrt{2}\sqrt{-\ol{\gamma}}} \dfrac{\delta S_{f,\tx{ren}}}{\delta \hat{\zeta}^{-}_{\ol{\phi}}} , \quad \exvalb{+}{\hzd{\phi}}\qft = \lim\limits_{z\rightarrow0} z^{-\frac{7}{2}}\hks^{-\frac{1}{12}}\, \exvalb{+}{\hzd{\phi}} ,\\
\lr{\bar{\mathcal{O}}^{+}_{\hat{\zeta}_{\hat{b}^\Phi}}} &=& \dfrac{i}{\sqrt{2}\sqrt{-\ol{\gamma}}} \dfrac{\delta S_{f,\tx{ren}}}{\delta \hat{\zeta}^{-}_{\ol{\bphi}}} , \quad \exvalb{+}{\hat{\zeta}_{\hat{b}^\Phi}}\qft = \lim\limits_{z\rightarrow0} z^{-\frac{7}{2}}\hks^{-\frac{1}{12}}\, \exvalb{+}{\hzd{\bphi}},
\end{split}
\eea
where $ S_{f,\tx{ren}} $ is the renormalized fermionic action.

\subsection{Diffeomorphisms and Local SUSY}

Our calculations in this sub-section and the next section are parallel to that given in \cite{Bertolini}, except for the fact that we have not explicitly introduced composite fields, but instead work with the diagonalized sources we have defined in \eqref{diagSources}. As mentioned in the previous subsection, our interest will remain with the bosonic sources $ \hat{\phi} $, $ \hat{b}^\Phi $, $ h_{\mu\nu} $ and their fermionic superpartners $ \hzd{\phi},\hzd{b} $ and $ \hat{\Psi}_{\mu} $

To get rid of the non-dynamical components in the metric and gravitino fields, we will choose the following gauge for the metric and gravitino 
\bea\label{gaugechoice}
ds^{2} =  dr^{2} + \gamma_{\mu\nu}(r,x) dx^{\mu}dx^{\nu}~,~~~~  \Psi_{r} = 0.
\eea
Since the calculations of the sources in the previous section are in a slightly more relaxed gauge (for metric and gravitino), we need to relate the two in explicit calculations. The metric ansatz chosen in \eqref{ktmetric} can be related via the identifications
$dr  = -\dfrac{e^{X}}{z}dz$, and  $\gamma_{\mu\nu} = \dfrac{e^{2Y}}{z^{2}} \eta_{\mu\nu}$.

In the remainder of this subsection we will look at bulk diffeomorphisms and supersymmetry transformations that preserve the gauge choice we have taken\footnote{As pointed out in \cite{Bertolini} and shown in \cite{Papadimitriou:2017kzw,An:2017ihs}, gauge-preserving bulk diffeomorphisms and local supersymmetry transformations generically mix. However the mixing involves transverse derivatives of the transformation parameters and are therefore subleading in the radial coordinate. Therefore this mixing will not affect our results and we can consider the two cases separately.}. In doing so, we will assume that the fields take the form in \eqref{asymptks}. That is, we will ignore the corrections that come at subleading orders to \eqref{asymptks}, with the understanding that in the asymptotic limit ($ z\rightarrow 0 $), where the QFT is defined, all the other contributions vanish sufficiently fast. 

\subsubsection{Weyl}

The set of bulk diffeomorphisms that preserve the gauge choice of the metric can be found by solving the Killing vector equations. The 5d Killing vector equations translate to
\bea
\partial_{r} \xi^{r} = 0, \ \ \  \partial_{r}\xi^{\mu } + \gamma^{\mu\nu} \partial_{\mu} \xi^{r} = 0.
\eea
The $\xi^{\mu}$ correspond to boundary diffeomorphisms and their Ward identities, which we are not interested in. The $\xi^r$ on the other hand can be interpreted as a Weyl transformation and is solved by $\xi^{r} = \sigma(x)$, which we will calculate. The action of the Weyl transformations on the bosonic fields is given by
\begin{align}
\begin{split}
\delta_{\sigma} \gamma_{\mu\nu} &= \sigma \partial_{r}\gamma_{\mu\nu} \ = \ - \sigma e^{-X}z \partial_{z}\gamma_{\mu\nu} \ = \ 2 \sigma \hks^{-\frac{2}{3}}\, \gamma_{\mu\nu} +\dots,\\
\delta_{\sigma} \phi &= - \sigma e^{-X}z \partial_{z}\phi \ = \ 0 +\dots ,\\
\delta_{\sigma} \bphi &= - \sigma e^{-X}z \partial_{z}\bphi \ = \ \gs q\sigma \hks^{-\frac{2}{3} } +\dots ~.
\end{split}
\end{align}
The variations of the hatted fields can be computed using the above
\bea
\des \ol{\phi} = 0 +\dots ~,~~~~\des \ol{\bphi} = \sigma \gs q \hks^{-\frac{2}{3}} +\dots ~,~~~~\des \ol{\gamma}_{\mu\nu} = 2\sigma \hks ^{-2/3}\ol{\gamma}_{\mu\nu}+\dots
\eea
For the fermionic fields, the action of the Weyl transformations is given by 
\begin{align}
\begin{split}
\des \zeta_{\phi}^{-} &= -\sigma e^{-X}z \de_{z}\zeta_{\phi}^{-} \ = \ -\dfrac{1}{2}\sigma\hks^{-\frac{2}{3}} \bigg(1+ \dfrac{1}{12\hks}\gs q^{2} \bigg)\zeta_{\phi}^{-} + \dots~,\\
\des \zeta_{b}^{-} &= -\sigma e^{-X}z \de_{z}\zeta_{b}^{-}~, \\
&=  -\dfrac{1}{2}\sigma\hks^{-\frac{2}{3}} \bigg(1+\dfrac{3}{8\hks} \gs q^{2}  \bigg)\zeta^{-}_{b} +\dfrac{1}{12}\sigma \hks^{-\frac{5}{3}} \Big(12-\dfrac{1}{\hks}\gs q^{2} \Big) \zeta^{-}_{\phi} + \dots,\\
\des \Psi_{\mu}^{+} &= -\sigma e^{-X}z \de_{z}\Psi^{+}_{\mu}\\
&= \dfrac{1}{2}\sigma\hks^{-\frac{2}{3}} \bigg(1+\dfrac{1}{12\hks} \gs q^{2}  \bigg)\Psi^{+}_{\mu} -\dfrac{i \gamma_{\mu}}{3}\sigma\bigg(q \hks^{-\frac{3}{2}} \Big[1+\dfrac{5}{24\hks}\gs q^{2}\Big]\zeta_{b}^{-}+\dfrac{5}{96\hks^{\frac{5}{2}}} \gs q^{4} \zeta_{\phi}^{-} \bigg) +\dots
\end{split}
\end{align}
We can use the above results and the definitions for the fermionic superpartners of the composite fields, to find their Weyl transformations. In the following, we have kept only the leading order terms of the powers of $\hks $ (the reason being $ \hks^{-\frac{2}{3}} |_{z\rightarrow 0} \sim \frac{1}{(\log z)^{2/3}} $ and anything subleading to $\hks^{-\frac{2}{3}}  $ will be falling off much faster, and will not contribute to the Ward identity computations)
\begin{align}
\begin{split}
\des \hzd{\phi}^{-} &= -\dfrac{1}{2}\sigma \,\hks^{-\frac{2}{3}}\,  \hzd{\phi}^{-}  + \dots~,\\
\des \hzd{b}^{-} &= -\dfrac{1}{2}\sigma \,\hks^{-\frac{2}{3}}\, \Big(\hzd{b}^{-} + \gs q \hzd{\phi}^{-} \Big) + \dots~,\\
\des \hat{\Psi}_{\mu}^{+} &= \dfrac{\sigma}{2}\hks^{-\frac{2}{3}}\hat{\Psi}_{\mu}^{+} + \dots~.
\end{split}
\end{align}

\subsubsection{Local Supersymmetry}

The fact that we have gauge-fixed the gravitino means that the gravitino SUSY variation in \eqref{fermionSUSYvar} gives rise to a differential equation for the supersymmetry parameter
\bea
\delta_{\epsilon}\Psi_{r} = \left(\nabla_{r} + \dfrac{1}{6}\calw \Gamma_{r} \right)\epsilon + \calo(z^{3}) = 0.
\eea
By projecting out the two chiralities (see discussions about spinors in the Appendix) with $ \Gamma_{r} \epsilon^{\pm} = \mp \epsilon^{\pm} $ and looking at the leading order terms in $ z $, we get
\bea
\partial_{r} \epsilon^{\pm} \mp \dfrac{1}{6}\calw \epsilon^{\pm} = 0, \ \ 
\Rightarrow \ \  \epsilon^{\pm}(z,x) = z^{\mp1/2}\hks(z)^{\pm1/12} \epsilon^{\pm}_{0} + \dots \ .
\eea
For the transverse coordinates of the gravitino
\bea
\delta_{\epsilon}\Psi_{\mu} =\nabla_{\mu}\epsilon + \dfrac{1}{6}\calw \Gamma_{\mu} \epsilon + \calo(z^{3}) 
= \partial_{\mu}\epsilon + \dfrac{1}{2} \omega_{\mu}^{zi}\, \gamma_{zi}\epsilon + \dfrac{e^{Y}}{6z} \calw \,  \delta_{\mu}^{i}\gamma_{i}\epsilon ++ \calo(z^{3})  .
\eea
Since we need only the leading asymptotics, we can project to the positive chirality of the gravitino, and using the on-shell values of $ \omega_{\mu}^{zi} $ and $ \calw $, we get
\bea
\delta_{\epsilon}\Psi_{\mu}^{+}&=& \partial_{\mu}\epsilon^{+} + \dfrac{1}{3} \calw \Gamma_{\mu} \epsilon^{-} + \calo(z^{3}) .
\eea
For the composite gravitino, using the above relation with \eqref{compgraviton} and \eqref{epsmu}, we get
\bea
\dep \hat{\Psi}_{\mu}^{+} = \hks^{-\frac{1}{6}}\de_{\mu}\epsilon^{+} + \hks^{-\frac{2}{3}} \hat{\Gamma}_{\mu} \epsilon^{-}+\dots
\eea

Now we turn to the SUSY-variations of the scalars and fermions. We can find the SUSY variations of the fermionic fields by evaluating to the relations given in \eqref{fermionSUSYvar} in the KS background. For the purpose of finding the SUSY Ward Identities, we need only the leading order results, given by
\begin{align}\label{epsmu}
\begin{split}
\depm \zeta_\phi^{-} &=  0 +\dots ~,~~~~~~\depm  \zeta_{b}^{-} = -i \gs q \hks^{-\frac{2}{3} }\,\epsilon^{-} +\dots  ~.\\
\end{split}
\end{align}
The supersymmetry transformations of the composite fields can be computed using the above results,
\bea
\depm \hzd{\phi}^{-} = 0 +\dots ~,~~~~~~\depm \hzd{b}^{-} =-i\gs q \hks^{-\frac{2}{3}} \,\epsilon^{-}+\dots ~.
\eea
For the bosonic fields of interest, we get the SUSY variations from \eqref{compferm} to be
\begin{align}
\begin{split}
\delta_{\epsilon^{+}} \hat{\phi} &=  \dfrac{i}{2}\Big( \bar{\epsilon}^{\, +} \hzd{\phi}^{-} - \hzdb{\phi} ^{\, -} \epsilon^{+} \Big) =  \dfrac{i}{2}\Big( \bar{\epsilon}^{\, +} \zeta_{\phi}^{-} - \bar{\zeta}_{\phi} ^{\, -} \epsilon^{+} \Big)\\
\delta_{\epsilon^{+}} \hat{b}^\Phi &= \dfrac{i}{2} \big( \bar{\epsilon}^{\, +}\hzd{b}^{-} -\hzdb{b}^{\, -} \epsilon^{+} \big) \ = \ \dfrac{i}{2} \big( \bar{\epsilon}^{\, +} \zeta_{b}^{-} - \bar{\zeta}_{b}^{-} \, \epsilon^{+}  \big) +  \dfrac{i}{2}\gs q \log az\big(\bar{\epsilon}^{\, +} \zeta_{\phi}^{-} - \bar{\zeta}_{\phi}^{-} \, \epsilon^{+} \big) +  \dots ,
\end{split}
\end{align}
where only the $ \epsilon^{+} $ variations are considered. This is because both $ \zeta_{\phi} $ and $ \zeta_{\bphi} $ are sourced by the negative chirality, and we are only interested in looking at the SUSY variations of the sources in this section.

We state this to emphasize the fact that we do not need an explicit form of the covariant sources as composite fields as was done in \cite{Bertolini}: we can derive all the necessary results we need in the computation of the 1-point functions using the above facts because only linear parts of variations show up in these calculations.

Finally, we turn to the SUSY variations of the metric. Using the supersymmetry transformation of the vielbein given in \eqref{bosonSUSYvar}, we can write the supersymmetry transformation of the boundary metric as
\bea
\begin{split}
\delta_{\epsilon}\gamma_{\mu\nu} &=& \delta_{\epsilon} \big(e_{\mu}^{a}e_{\nu}^{b}\eta_{ab} \big) = \dfrac{1}{2} \Big(\bar{\epsilon} \, \Gamma_{\mu} \Psi_{\nu} + \bar{\epsilon} \, \Gamma_{\nu}\Psi_{\mu} \Big) + \text{h.c.}  \\
&=& \bar{\epsilon}^{\,+} \, \Gamma_{(\mu} \Psi_{\nu)}^{+} +\bar{\epsilon}^{\, -} \, \Gamma_{(\mu} \Psi_{\nu)}^{-}+ \text{h.c.}.
\end{split}
\eea
The symmetrization here contains the factor of $ 1/2 $. We can drop the $ \Psi_{\nu}^{-} $ owing to the fact that it is subleading (and therefore does not corresponds to a source), and we get
\bea
\begin{split}
\delta_{\epsilon^{+}}\gamma_{\mu\nu} &=& \bar{\epsilon}^{+} \, \Gamma_{(\mu} \Psi_{\nu)}^{+} + \text{h.c.} + \dots,\\
\Rightarrow \delta_{\epsilon^{+}} \ol{\gamma}_{\mu\nu} &=&  \bar{\epsilon}^{+} \, \hat{\Gamma}_{(\mu} \hat{\Psi}_{\nu)}^{+} + \text{h.c.} + \dots
\end{split}
\eea

\subsection{Derivation of SUSY and Trace Ward Identities}\label{sec5.2}

Now we can put all these ingredients together to compute the SUSY and trace Ward identities. We directly present the results: the approach is parallel to that in \cite{Bertolini}. In order to compute the SUSY Ward identities, we only turn on those sources that do not break SUSY explicitly. As noted in the description of the linearized sources for the bosons, the relevant sources that are present in the \zz theory explicitly break SUSY. Hence, to compute the SUSY Ward Identities, we take the action to be a functional of the SUSY preserving sources, 
\bea
S_{ren} \equiv S_{ren} \Big[\hat{\gamma}_{\mu\nu},\hat{\phi},\hat{b}^\Phi,\hat{\Psi}_{\mu},\hzd{\phi}^{-},\hzd{b}^{-} \Big]~.
\eea
The action we use to compute the Ward Identities is to be understood as the full $ \caln =2 $ renormalized supergravity action, with both the bosonic and fermionic fields. However, we do not need the explicit form of the action to carry out the computations.

In computing the SUSY Ward/Trace Identities, the set of sources used here are the one that appear in the $ U(1) $ truncation as well. However, the presence of more fields does change the SUSY variations of substantially. The reason we go through this section (and Appendix \ref{appd}) in such detail is to ensure that all the falloffs that go in to the Ward/Trace identity computations are under control.
\subsubsection{SUSY Ward Identities}

We consider $\epsilon^{+}, \epsilon^{-}$ and $\sigma$ in turn. We use the results from the previous sub-section, where we found the action of $ \sigma,\epsilon^{\pm} $ on the sources, to compute the Ward Identities. First, we will look at the $ \epsilon^{+} $ variation which will give rise to SUSY Ward identities in the boundary QFT. We have
\begin{align}
\delta_{\epsilon^{+} }S_{\tx{ren}} &= \int d^{4}x\sqrt{-\ol{\gamma}} \Big(\dfrac{i}{2} \lr{\bar{S}^{\mu-}}\delta_{\epsilon^{+}}\hat{\Psi}^{+}_{\mu} + \dfrac{1}{2}\lr{T^{\mu\nu}} \delta_{\epsilon^{+}}\ol{\gamma}_{\mu\nu} +2 \exval{+} \delta_{\epsilon^{+}}\ol{\phi}  + 2 \exval{-} \delta_{\epsilon^{+}}\ol{\bphi}  +\text{h.c.}\Big)\non \\
&= \int d^{4}x\sqrt{-\ol{\gamma}} \bigg[ -\dfrac{i}{2}\lr{\de_{\mu} \bar{S}^{\mu-}}\hks^{-\frac{2}{3}} -\dfrac{1}{2} \lr{T^{\mu\nu} } \hat{\Gamma}_{\mu}\hat{\Psi}^{+}_{\mu} -i\exval{+} \hzdb{\phi}^{-} -i\exval{-} \hzdb{b}^{-} \bigg]\epsilon^{+} +\text{h.c.} 
\end{align}
In these formulas (and formulas in subsequent subsections), we have used \eqref{foneptfundef} for the definition of fermionic one-point functions at non-zero source. For one-point function of the stress-tensor and other bosonic operators, we use the definition in \eqref{oneptfn1}. By setting $ \delta_{\epsilon^{+}}S = 0 $, we get the following operator relation at a finite radial cut-off surface at non-zero sources
\bea
\dfrac{i}{2}\hks^{-\frac{1}{6}}\, \lr{\de_{\mu} \bar{S}^{\mu-}}  &=& -  \dfrac{1}{2} \lr{T^{\mu\nu} } \overline{\hat{\Psi}}^{+}_{\mu}\hat{\Gamma}_{\nu} -i\exval{+} \hzdb{\phi} -i\exval{-} \hzdb{b}^{-} 
\eea
Taking the functional derivatives of this w.r.t. the different fermionic sources gives rise to the following identities
\begin{subequations}\label{SUSYWI}
\bea
\hks^{-\frac{1}{6}}\,\lr{\de_{\mu} \bar{S}^{\mu-}(x)\bar{S}^{\nu-}(0)}  &=& 2i\hat{\Gamma}_{\mu}\lr{T^{\mu\nu}}\delta^{4}(x,0)~,\\
\hks^{-\frac{1}{6}}\,\lr{\de_{\mu} \bar{S}^{\mu-}(x)\calo^{+}_{\hzd{\phi}}(0)}  &=& \sqrt{2}\exval{+}\delta^{4}(x,0)~,\\
\hks^{-\frac{1}{6}}\,\lr{\de_{\mu} \bar{S}^{\mu-}(x)\calo^{+}_{\hzd{b}}(0)}  &=& \sqrt{2}\exval{-}\delta^{4}(x,0)~,
\eea
\end{subequations}
where $ \delta^{4}(x,y) = \sqrt{-\ol{\gamma}}\,\delta^{4}(x-y) $ . Now, using the definitions for the QFT one-point functions, we finally get SUSY Ward identities
\begin{subequations}
\bea\label{sWI}
\lr{\de_{\mu} \bar{S}^{\mu-}(x)\bar{S}^{\nu-}(0)}\qft  &=& 2i\gamma_{i}\delta^{i}_{\mu}\lr{T^{\mu\nu}}\qft \delta^{4}(x),\\
\label{OpWI}
\lr{\de_{\mu} \bar{S}^{\mu-}(x)\calo^{+}_{\hzd{\phi}}(0)}\qft  &=& \sqrt{2}\exval{+}\qft\delta^{4}(x),\\
\label{OmWI}
\lr{\de_{\mu} \bar{S}^{\mu-}(x)\calo^{+}_{\hzd{b}}(0)}\qft  &=& \sqrt{2}\exval{-}\qft\delta^{4}(x).
\eea
\end{subequations}
Eqn. \eqref{sWI}, is the Ward identity involving, the supercurrent and the stress tensor, which sit in the supercurrent multiplet. Eqns. \mref{OpWI, OmWI} are Ward identities for operator sitting in a chiral supermultiplet in which the highest component operators are ${\cal O}_{\pm}$. Therefore a non-zero vev for ${\cal O}_{\pm}$ would correspond to a supersymmetry broken vacuum. However, this is not the sole criterion for spontaneous supersymmetry breaking, as we will see in the next subsection.
\subsubsection{Trace Identities}
Looking at the variation of $S_{\tx{ren}}$ under $\epsilon^{-}$ gives us
\bea
\begin{split}
\delta_{\epsilon^{-}} S_{\tx{ren}}  &=& \int d^{4}x \sqrt{-\ol{\gamma}} \, \bigg[ \dfrac{i}{2}\lr{\bar{S}^{\mu-}} \delta_{\epsilon^{-}}\hat{\Psi}_{\mu}^{+} -i\sqrt{2} \exvalb{+}{\hzd{b}}  \delta_{\epsilon^{-}}\hzd{b}^{-}  +\text{h.c}  \bigg]~,\\
&=& \int d^{4}x \sqrt{-\ol{\gamma}} \, \bigg[\dfrac{i}{2}\hks^{-2/3}\lr{\bar{S}^{\mu-}} \hat{\Gamma}_{\mu} -\sqrt{2} \gs q\, \hks^{-2/3} \exvalb{+}{\hzd{b}} \bigg]\epsilon^{-} +\tx{h.c.}
\end{split}
\eea
Setting $ \delta_{\epsilon^{-}}S_{\tx{ren}} = 0 $, we get
\bea
\dfrac{i}{2}\lr{\bar{S}^{\mu-}} \hat{\Gamma}_{\mu} &=& \sqrt{2}  \gs q\exvalb{+}{\hzd{b}} ~.
\eea
Finally, taking the boundary limit, we get the following QFT operatorial relation between the $\gamma$-trace of the supercurrent and the fermionic superpartner of ${\cal O} _-$ (upto potential anomaly terms\footnote{To calculate the anomaly terms we need to know the explicit form of the bosonic and fermionic counterterm. For a systematic derivation of these terms in $4d$, ${\cal N}=1$ and $3d$, ${\cal N}=2$ superconformal theories on an arbitrary curved background, see \cite{Papadimitriou:2017kzw,Papadimitriou:2004rz,An:2017ihs}.}).
\bea\label{superWeylWI}
\dfrac{i}{2}\lr{\bar{S}^{\mu-} \gamma_{i}\delta^{i}_{\mu}}\qft &=& \sqrt{2} \gs q\exvalb{+}{\hzd{b}}\qft~.
\eea
Finally we consider the invariance of the $S_{\tx{ren}}$ under rescalings of the radial coordinate. We have
\bea
\des S_{\tx{ren}}  &=& \int d^{4x}\sqrt{-\ol{\gamma}}\bigg[ \dfrac{1}{2}\lr{T^{\mu\nu}}\des \ol{\gamma}_{\mu\nu} + 2\exval{-} \des \ol{\bphi}  +\dfrac{i}{2}\lr{\bar{S}^{\mu-}}\des \hat{\Psi}_{\mu}^{+} \non\\
&& \qquad\qquad\qquad -\sqrt{2}i\Big(\exvalb{+}{\hzd{\phi}}\des \hzd{\phi}^{-} +\exvalb{+}{\hzd{b}}\des \hzd{b}^{-} \Big)+ \text{h.c.}\bigg]~.
\eea
Using the $ \sigma $-variations and turning off the fermionic sources, we get
\bea
\lr{T^{\mu}_{\mu}} &=& -2 \gs q ~\exval{-}~.
\eea
By taking the boundary limit, we get the following relation between the trace of the stress-tensor and the operator ${\cal O}_-$
\bea\label{WeylWI}
\lr{T^{\mu}_{\mu}}\qft &=& -2 \gs q~ \exval{-}\qft ~ ,
\eea
upto potential anomaly terms which do not appear here since we have taken the boundary metric to be Minkowskian. The relation \eqref{WeylWI} is the bosonic counterpart of the fermionic relation in \eqref{superWeylWI} and the two results are in perfect agreement. In the next section we will check these Ward identities  on a vacua of the KS theory dual to the two-parameter SUSY breaking solution in \eqref{KSSUSYb} by explicitly calculating the one-point functions. This will allows us to comment upon the nature of supersymmetry breaking.

\section{One-point Functions and the Goldstino Pole}\label{sec:6}

To obtain the QFT one point functions, we evaluate the functional derivatives of the renormalized on-shell action appearing \eqref{oneptfn2} and take the limits in \eqref{oneptfn3}. The regulated action in $S_{\tx{reg}}$ is given by
\bea
S_{\tx{reg}} = S_{\tx{5D}} + S_{\tx{GH}}~,
\eea
where $S_{\tx{5D}}$ the boundary contribution coming from the five-dimensional gauged supergravity action and $S_{\tx{GH}}$ is the Gibbons-Hawking term. Correlation functions computed from $S_{\tx{reg}}$ are typically divergent because of the infinite volume of spacetime. Finite quantities can be obtained through the standard procedure of holographic renormalization where we first identify the divergences of the regularized on-shell action and then add appropriate local covariant counterterms to kill these divergence \cite{deHaro,Papadimitriou:2004ap,Avinash}. The renormalized action thus obtained is finite when the cut-off surface is taken to the boundary. However there are scheme ambiguities associated to finite terms which may be required to preserve supersymmetry. For flat-domain wall solution the superpotential ${\cal W}$ in \eqref{superpotential} has all the necessary finite terms to render $ S_{\tx{ren}}=S_{\tx{reg}}+S_{\tx{ct}} =0 $ on supersymmetric configurations \cite{Bianchi}. Therefore we take the following as out bosonic counter term 
\bea\label{counterterm}
S_{\tx{ct}} = -\int d^{4}x\sqrt{-\gamma } ~2\calw~.
\eea
This, along with the fact that counter terms have to be universal for any solution to the equations of motion for the given potential, fixes them once and for all, regardless of the bulk solution being supersymmetric or not. The calculation of the one point functions for the marginal operators proceed as in \cite{Bertolini} and we find no further subtleties.
\begin{subequations}
\begin{align}
\label{optT}
\lag T^\m_{~\n}\rag_{\tx{QFT}}&= -3 {\cal S}a^4 \delta^\m_{~\n}~,\\
\label{optOp}
\lag {\cal O}_+\rag_{\tx{QFT}}&= \frac12\(3{\cal S}+4\varphi\)a^4~,\\
\label{optOm}
\lag {\cal O}_-\rag_{\tx{QFT}}&= \frac{6{\cal S}}{g_s q}a^4~,
\end{align}
These expressions were first obtained in \cite{DKM} and were later independently derived in \cite{Bertolini}. Here we find that even in the full KS theory, these one-point functions remain unaffected (upto a trivial modification by the conifold deformation parameter $a$ defined in \eqref{warpKS}). 

We now expand around this result. Integrating the SUSY ward identity in \eqref{sWI} in $x_\mu$ gives us the two point function of the supercurrent\cite{Argurio,Argurio0}. The right hand side contains a massless fermionic pole provided the vev of the stress-tensor (that corresponds to the vacuum of a QFT state) is non-zero. This massless pole is hallmark signature of the presence of a goldstino which is associated to the spontaneous breaking of supersymmetry. Since from \eqref{optT} we have that the one-point function of the stress-tensor is non-zero and gets contribution from the parameter ${\cal S}$ only, we conclude that ${\cal S}$ corresponds to spontaneous supersymmetry breaking. Furthermore, the one-point functions in \mref{optT,optOm} satisfy the trace identity derived in \eqref{WeylWI}. We see that there is no contribution to the vacuum energy from the vev of ${\cal O}_+$. On the other hand the parameter $\varphi$ corresponds to explicit breaking of supersymmetry and does not corresponds to a vacua of the KS gauge theory. This is because in this SUSY breaking solution (where ${\cal S}=0$), the vacuum energy vanishes ($\lag T^\mu_{~\mu}\rag=0$) and therefore the residue of the Goldstino pole vanishes. Despite the technical complications, we find that even in the full theory, which captures the conifold deformation parameter, all of the aforementioned results are identical to those obtained via the $U(1)$-truncated supergravity action considered in \cite{DKM,Bertolini}.

Finally let us comment upon the one-point functions of the gaugino bilinear operators ${\cal Q}_\pm$ which do not participate in any of the Ward identities (since we have not turned on sources for these operators). Without the inclusion of the counterterm in \eqref{counterterm} we find that the vev of ${\cal Q}_+$ is finite and its value is 
\begin{align}
\lag {\cal Q}_+\rag_{\tx{QFT}}&= \dfrac{18 a^3}{\gs q}~.
\end{align}
Hence the one-point function of ${\cal Q}_+$ does not require renormalization. The counterterm gives the same finite contribution but with opposite sign. Therefore if we include the counterterm contribution we will not be able see the vev which is actually non-zero. The inverse dependence of the vev on $q$ can be attributed to the normalization of the operator ${\cal Q}_+$. In \cite{Loewy:2001pq} the operator ${\cal Q}_+$ (as defined in \eqref{sourceOp}) was identified with the gaugino bilinear that condense in the Klebanov-Strassler gauge theory. Here we find that even in the presence of the SUSY breaking perturbations, this vev is uneffected. On the other hand the vev of ${\cal Q}_-$ is divergent, and therefore needs renormalization. Upon adding the counterterm \eqref{counterterm}, we do not find any non-vanishing finite contributions. Therefore we conclude that 
\begin{align}
\lag {\cal Q}_-\rag_{\tx{QFT}}&= 0~.
\end{align}
\end{subequations}

It would be interesting to generalize our analysis to the full $SU(2) \times SU(2)\times \Z_2$ truncation by turning on fields in \eqref{nonKSfields} to see if there are more SUSY-breaking parameters and study if there exists spontaneously supersymmetry breaking vacua. Our minimal goal here, namely to derive the SUSY Ward identities in a truncation of Type IIB SUGRA, that admits the deformation of the conifold parameter, has been accomplished. Since the parameter $\mathcal{S}$ is known to be triggered by anti-D3 branes on the tip of the throat \cite{DKM}, this shows that {\em if} the KKLT construction is (meta-) stable\footnote{See discussions on some aspects of this issue in \cite{Sethi, Halmagyi, Polchinski, Dasgupta}}, it is a spontaneously broken (and therefore bonafide) vacuum of string theory. The goldstino on the worldvolume of the anti-D3 brane has been noted in previous work in \cite{Wrase}. 

\section*{Acknowledgments}

We thank K. V. Pavan Kumar for collaboration at the early stages of this project and Matteo Bertolini, Daniele Musso,  Avinash Raju and Aninda Sinha for discussions. CK thanks Phil Szepietowski for a helpful correspondence regarding \cite{Liu:2011dw} at the early stages of this work. We thank Justin David for extending warm hospitality to one of the authors (HR) at IISc during the very final stages of this project. HR gratefully acknowledges the warm hospitality and support from the Mani L. Bhaumik Institute for Theoretical Physics. 

\appendix

\section{Truncations, Ansatzes and Uplifts}\label{appa}

The KT solution in the Type IIB setting and the linearized SUSY breaking perturbations that asymptote to KT were discussed in \cite{DKM} and in terms of 5d Supergravity in \cite{Bertolini} (where they use the notations of \cite{Buchel,Cassani}). We will discuss some salient points in the uplift of 5d Supergravity solutions to the 10d Type IIB. This will serve to both establish the correspondence with the notations in various previous papers, as well as to emphasize some subtleties.

In the notation of \cite{DKM}, the 10d metric for the $ U(1) $ truncation is given by
\bea\label{DKMmetric}
ds^{2}_{10\,(DKM)} = r^{2}e^{2a(r)}\eta_{\mu\nu}dx^{\mu}dx^{\nu} + \dfrac{e^{-2a(r)}}{r^{2}} dr^{2} +  \dfrac{1}{6}e^{2(c(r)-a(r))}\sum_{a=1}^{4} (\ee^{a})^{2} + \dfrac{1}{9}e^{2(b(r)-a(r))} (\ee^{5})^{2}~.\non\\
\eea
The two scalar fields coming from the dilaton and B-field of IIB are denoted by $ \Phi(r) $ and $ k(r) $ in \cite{DKM}, which are denoted by $ \phi(z) $ and $ \bphi(z) $, respectively, in our paper, and the radial coordinates are related as $ r=1/z $. The linearized solution to the equations of motion around the KT background allows for perturbations of the fields $ \{a,b,c,k,\Phi\} $. 

The 10d metric in the notation of \cite{Buchel}, keeping only the fields corresponding to the $ U(1) $ truncation is given by
\bea
ds_{10}^{2} = e^{-\frac{2}{3}(4u+v)} ds^{2}_{5} + \dfrac{1}{6} e^{2u} \sum_{a=1}^{4} (\ee^{a})^{2} + \dfrac{1}{9}e^{2v} (\ee^{5})^{2},
\eea
where $ ds^{2}_{5} = g_{AB} dx^{A}dx^{B} $ is the 5d metric. In \cite{Bertolini} the 5d metric is taken to be of the form
\bea
ds^{2}_{5} = \dfrac{1}{z^{2}} \left(e^{2X} dz^{2} + e^{2Y} \eta_{\mu\nu} dx^{\mu} dx^{\nu}\right).
\eea
The equations of motion are solved using the parametrization
\bea
e^{X(z) } = h(z)^{\frac{2}{3}}\,h_{2}(z)^{\frac{1}{4}},&& \ e^{Y(z)} = h(z)^{\frac{1}{6}}\, h_{2}(z)^{\frac{1}{4}}\,h_{3}(z)^{\frac{1}{4}} \non\\
e^{U(z)} = h(z)^{\frac{5}{4}} \, h_{2}(z)^{\frac{3}{4}}, && \ e^{V(z)} = h_{2}(z)^{-\frac{3}{4}},
\eea
where $ U=4u+v $ and $ V=u-v $. On uplifting this ansatz to 10d, this is in a slightly different gauge for the radial coordinate comapred to \cite{DKM}: 
\bea
ds_{10}^{2} = \dfrac{h(z)^{-\frac{1}{2}}h_{3}(z)^{\frac{1}{2}}}{z^{2}} \eta_{\mu\nu}dx^{\mu}dx^{\nu} + \dfrac{h(z)^{\frac{1}{2}}}{z^{2}} dz^{2} +  \dfrac{1}{6} h(z)^{\frac{1}{2}}\sum_{a=1}^{4} (\ee^{a})^{2} + \dfrac{1}{9}h(z)^{\frac{1}{2}} h_{2}(z)^{\frac{3}{2}} (\ee^{5})^{2}.
\eea
After a coordinate change to $ r=1/z $ and defining $ H^{\frac{1}{2}} = r^{-2} h^{\frac{1}{2}} h_{3}^{-\frac{1}{2}} $, we get
\bea
ds_{10}^{2} = H^{-\frac{1}{2}} \eta_{\mu\nu}dx^{\mu}dx^{\nu} + H^{\frac{1}{2}} \bigg( h_{3}^{\frac{1}{2}}dr^{2} +  \dfrac{1}{6} r^{2}h_{3}^{\frac{1}{2}}\sum_{a=1}^{4} (\ee^{a})^{2} + \dfrac{1}{9}r^{2}h_{3}^{\frac{1}{2}} h_{2}^{\frac{3}{2}} (\ee^{5})^{2}\bigg)
\eea

The most general parametrization of the functions can be taken in the form
\bea
e^{X(z) } =h_{X}^{\frac{2}{3}}(z),&& \ e^{Y(z)} =  h_{Y}^{\frac{1}{6}}(z)\non\\
e^{U(z)} =  h_{U}^{\frac{5}{4}}(z), && \ e^{V(z)} = h_{V}(z).
\eea
where $h_{X},\, h_{Y} $ and $ h_{U} $ at leading order is given by $ \hks $ and $ h_{V} =1 $ at leading order. The  functions are each a double series in $ z^{n} $ and $ z^{n}\log z $.This metric uplifts to
\bea
ds_{10}^{2} &=& \dfrac{h_{U}^{-\frac{5}{6}}\, h_{Y}^{\frac{1}{3}}}{z^{2}}\eta_{\mu\nu}dx^{\mu}dx^{\nu} + \dfrac{ h_{U}^{-\frac{5}{6}}\, h_{X}^{\frac{4}{3}}}{z^{2}} dz^{2} \non\\
&& +  \dfrac{1}{6} h_{U}^{\frac{1}{2}}\, h_{V}^{\frac{2}{5}}\sum_{a=1}^{4} (\ee^{a})^{2} + \dfrac{1}{9}  h_{U}^{\frac{1}{2}}\, h_{V}^{-\frac{8}{5}} (\ee^{5})^{2}.
\eea
The equations of motion can be solved order by order for this ansatz (we also include the other fields in the $\IZ_2$ truncation to do this, obviously), and we find that there are a total of 4 independent (SUSY-preserving) parameters on top of the SUSY-breaking ones. 

Note however that the above ansatz is not the most convenient for a few reasons. Firstly, we have not fixed the gauge freedom (this in particular means that we cannot be sure that all the perturbations we found are physical), and secondly, we find it (slightly) better to work with an ansatz that is more naturally adapted to a 10d brane ansatz form in the spirit of \cite{DKM}. A (partial) gauge fixing that accomplishes this is the ansatz we use in the main body of the paper:
\bea
e^{X(z) } = h(z)^{\frac{2}{3}}\,h_{2}(z)^{\frac{1}{3}}h_{3}(z)^{\frac{4}{3}},&& \ e^{Y(z)} = h(z)^{\frac{1}{6}}\,h_{2}(z)^{\frac{1}{3}}h_{3}(z)^{\frac{4}{3}}  \non\\
e^{U(z)} = h(z)^{\frac{5}{4}} \, h_{2}(z)\, h_{3}(z)^{4}, && \ e^{V(z)} = h_{3}(z)\, h_{2}(z)^{-1}, \label{ouransatz}
\eea
which in the $U(1)$ case, when uplifted to 10d takes the form
\bea
ds_{10}^{2} = \dfrac{h(z)^{-\frac{1}{2}}}{z^{2}}\eta_{\mu\nu}dx^{\mu}dx^{\nu} + \dfrac{h(z)^{\frac{1}{2}}}{z^{2}} dz^{2} +  \dfrac{1}{6} h(z)^{\frac{1}{2}}h_{3}(z)^{2}\sum_{a=1}^{4} (\ee^{a})^{2} + \dfrac{1}{9}h(z)^{\frac{1}{2}}h_{2}(z)^{2} (\ee^{5})^{2}.
\eea
It is straightforward to see that this metric and the metric in \eqref{DKMmetric} are the same form upto renaming of functions, with the identification $ r = 1/z $. For the \zz truncation, the same ansatz lifts to a metric of the form
\bea
ds_{10}^{2} &=& \dfrac{h(z)^{-\frac{1}{2}}}{z^{2}}\eta_{\mu\nu}dx^{\mu}dx^{\nu} + \dfrac{h(z)^{\frac{1}{2}}}{z^{2}} dz^{2} +  \dfrac{\cosh t}{6} h(z)^{\frac{1}{2}}h_{3}(z)^{2}\sum_{a=1}^{4} (\ee^{a})^{2} \non\\
&& \qquad + \dfrac{\sinh t}{3}h(z)^{\frac{1}{2}}h_{3}(z)^{2} \Big( \ee^{1} \ee^{3} + \ee^{2} \ee^{4} \Big)  + \dfrac{1}{9} h(z)^{\frac{1}{2}}h_{2}(z)^{2} (\ee^{5})^{2}.
\eea
The advantage of this ansatz, which is the one we use in this paper, is that it removes all the SUSY-preserving perturbations except for one (which we argue in the next Appendix is a gauge mode).
\section{Gauge Freedom in the 10d metric}\label{appb}

In this appendix, we will show that a specific perturbation that arises in the class of 10d metrics from the previous section when expanded around Klebnov-Witten, is a coordinate redefinition. The reason for our interest in this perturbation is that within the ansatzes that we work with\footnote{By which we mean the forms \eqref{ouransatz} as well as the combined expansions in $z^n$ and $z^n \ln z$ with $n$ not restricted to be even. If $z$ is restricted to be even as in \cite{Aharony} this term does not arise and this appendix can be skipped.}, this is the only perturbation (SUSY-preserving) that shows up around the KS background other than the parameters in KS and the SUSY-breaking perturbations. The fact that precisely this perturbation arises also around KW, and there it can be understood as a gauge artefact will be taken as motivation to believe that it is a gauge artefact around KS as well. We will work with the $U(1)$ truncation to keep the notation slightly cleaner, but the arguments go through precisely analogously in the $\IZ_2$ case as well.

Let us start with the 10d metric 
\bea
ds_{10}^{2} = \dfrac{h(z)^{-\frac{1}{2}}}{z^{2}}\eta_{\mu\nu}dx^{\mu}dx^{\nu} + \dfrac{h(z)^{\frac{1}{2}}}{z^{2}} dz^{2} +  \dfrac{1}{6} h(z)^{\frac{1}{2}}h_{3}(z)^{2}\sum_{a=1}^{4} (\ee^{a})^{2} + \dfrac{1}{9}h(z)^{\frac{1}{2}}h_{2}(z)^{2} (\ee^{5})^{2}.
\eea
The KW solution is given by $ h(z) = h_{2}(z) = h_{3}(z)=1 $. Now, let us look at small arbitrary perturbation around this background. The metric becomes
\bea
ds_{10}^{2} &=& \dfrac{ (1+\delta  h(z))^{-\frac{1}{2}}}{z^{2}} \eta_{\mu\nu}dx^{\mu}dx^{\nu} + \dfrac{ (1+\delta  h(z))^{\frac{1}{2}}}{z^{2}} dz^{2}  +  \dfrac{1}{6}(1+\delta  h(z))^{\frac{1}{2}} (1+\delta  h_{3}(z))^{2}\sum_{a=1}^{4} (\ee^{a})^{2} \non\\
&& \qquad\qquad + \dfrac{1}{9}(1+\delta  h(z))^{\frac{1}{2}} (1+\delta  h_{2}(z))^{2} (\ee^{5})^{2}.
\eea
We can redefine the $ z $-coordinate to $ y $ in the following way
\bea
\dfrac{z^{2}}{ (1+\delta  h(z))^{-\frac{1}{2}}} &=& y^{2}\qquad \Rightarrow\quad y^{2} \ \simeq \  z^{2} \Big(1+\dfrac{1}{2}\delta h(z)\Big) \\
2y dy &=& \Big[2z\Big(1+\dfrac{1}{2}\delta h(z)\Big) +\dfrac{z^{2}}{2} \delta h'(z) \Big] dz.
\eea
We will only need the perturbation upto linear order, so these approximations will turn out to be consistent for our purposes.
Using this we get
\bea
\dfrac{ (1+\delta  h(z))^{\frac{1}{2}}}{z^{2}} dz^{2} &\approx& \dfrac{4y^{2}\Big(1+ \frac{1}{2}\delta h(z)\Big) dy^{2} }{y^{2}\Big(1-\frac{1}{2}\delta h(z)\Big) \Big[2z\Big(1+\frac{1}{2}\delta h(z)\Big) +\frac{z^{2}}{2} \delta h'(z) \Big]^2}\non\\
&\approx& \dfrac{4 dy^{2}}{\Big(1-\delta h(z)\Big) \Big( 4 z^{2}(1+ \delta h(z)) + 2 z^{3}  \delta h'(z) \Big)}\non\\
&\approx& \dfrac{dy^{2}}{z^{2}\Big(1+\frac{1}{2}z\, \delta h'(z)\Big)}.
\eea
If we now set $ \delta h(z) = \e z $, the denominator in the last line can be rewritten as $ z^{2}\Big(1+ \frac{1}{2} \e z\Big) = z^{2}\Big(1+ \frac{1}{2} \delta h(z)\Big) \simeq y^{2} $. Thus, we get
\bea
\dfrac{ (1+\delta  h(z))^{\frac{1}{2}}}{z^{2}} dz^{2} &\simeq& \dfrac{dy^{2}}{y^{2}}. 
\eea
In order to have the full metric unchanged under this redefinition, we need
\bea
(1+\delta  h(z))^{\frac{1}{2}} (1+\delta  h_{2}(z))^{2} = (1+\delta  h(z))^{\frac{1}{2}} (1+\delta  h_{3}(z))^{2} =1.
\eea
Altogether these conditions read
\bea
\delta h_{2}(z) = \delta h_{3}(z) = - \dfrac{1}{4} \delta h(z)=  -\dfrac{1}{4}\e z.
\eea

The reason we care about this, is because the 10d metric we started with, when expanded around KW has precisely this as a perturbation at $\mathcal{O}(z)$ when we demand that the equations of motion hold. This means that that particular perturbation can be viewed as a gauge artefact.

\section{Fermions in AdS: A mini review}

The spin-1/2 fermions in 5d are Dirac fermions. The gamma matrices are given by
\bea
\Gamma^{A} &=& e^{A}_{a}\gamma^{a},
\eea
where $ e^{A}_{a} $ are the vielbeins corresponding to the 5d metric. The $ \gamma^{a} $'s can be grouped into the gamma matrices of the boundary 4-d space $ \gamma^{i}, i=0,1,2,3 $ and $ \gamma^{z} $ of the radial direction
\bea
\{\gamma^{i},\gamma^{j}\} &=& 2\eta^{ij}, \ \ \ \gamma^{0\dagger} \ = \ -\gamma^{0}, \ \ \ \gamma^{i \dagger} \ = \ \gamma^{0}\gamma^{i}\gamma^{0},\\
\{\gamma^{z},\gamma^{j}\} &=& 0, \ \ \ \ \ \gamma^{z}{}^{2} \ = \  1,  \ \ \ \gamma^{z\dagger} \ = \ \gamma^{z}.
\eea
The conjugate spinor is defined as
\bea
\overline{\psi} &=& \psi^{\dagger}\,i \gamma^{0}.
\eea

One basic idea in solving fermionic fields in AdS is that a spinor in the bulk, being a 5d spinor has the same number of components as a 4D Dirac spinor. But the minimal spinors on the boundary are (4D) Weyl spinors and contain half as many degrees of freedom. When we want to use them as boundary data for solving the bulk (spinor) equations, the two possible chirality choices separate out. This is good: because unlike in the bosonic cases, the bulk spinor equations are first order. So it is good that the two chiralities on the boundary can yield a natural interpretation as source and condensate - as they do in the bosonic case for the field and its derivative (roughly).

Lets see how this works out in the case of Rarita-Schwinger fields and spin-1/2 fermions. The latter discussion we follow the very clear presentation in \cite{Henneaux}.

\subsection{Rarita-Schwinger field in AdS}

The Rarita-Schwinger equation, in the AdS background, for a gravitino of mass $ m=\frac{3}{2} $, is given by
\bea \label{rsrhoeq}
(\delta^{\rho}_{j}\delta^{\mu}_{i}\gamma^{j}\gamma^{i} - \eta^{\rho\mu})\bigg(-z^{3}\gamma^{z}\,\de_{z} \Psi_{\mu}(z,x) + \dfrac{z^{2}}{2}  (2\gamma^{z}-3) \Psi_{\mu}(z,x)\bigg)  &&\non\\
+z^{3}(\delta^{\rho}_{j}\delta^{\nu}_{k}\delta^{\mu}_{i}\gamma^{j} \gamma^{k}\gamma^{i} - \eta^{\rho\nu} \delta^{\mu}_{i}\gamma^{i} -  \eta^{\mu\nu} \delta^{\rho}_{j}\gamma^{j} + \eta^{\rho\mu} \delta^{\nu}_{k}\gamma^{k} )\,\de_{\nu} \Psi_{\mu}(z,x) \ = \ 0 ,&&\\
\Rightarrow z^{3} (\delta^{\nu}_{j}\delta^{\mu}_{i} \gamma^{j}\gamma^{i}- \eta^{\mu\nu}) \de_{\nu}\Psi_{\mu}(z,x) + \label{rszeq}\dfrac{3 z^{2}}{2} \delta^{\mu}_{i}\gamma^{i} (1-\gamma^{z}) \Psi_{\mu}(z,x) \ = \ 0.&&
\eea
We can solve these equations near the boundary $ z=0 $, using the Frobenius method. We substitute the series expansion
\bea
\Psi_{\mu}(z,x) = z^{\Delta}\sum_{l=0} c_{\mu(i)}(x) z^{l},
\eea
in \eqref{rsrhoeq}, and set the coefficients of each of the $ z $ powers to zero. The leading equation is 
\bea
\Big(-\Delta \gamma^{z} +\gamma^{z} -\dfrac{3}{2} \Big) c_{\mu(0)}(x) = 0,
\eea
which is solved by
\bea
\Delta = \begin{cases}
	-\frac{1}{2} \qquad &\text{with}\ \ \gamma^{z}c_{\mu(0)}(x) = c_{\mu(0)}(x) \\
	\frac{5}{2} \qquad &\text{with}\ \ \gamma^{z}c_{\mu(0)}(x) = -c_{\mu(0)}(x).
\end{cases}
\eea
These two are the two independent boundary fields that fix the full gravitino solution in the bulk.

We stress here that this discussion is for the AdS background and not the KS background, where the fermions and gravitino are non-trivially coupled. 

\subsection{Spinors in AdS}

For simplicity, we will consider a single fermion of mass $ m $ in the AdS background. The equation of motion for the fermion is 
\bea\label{fermads}
z \gamma^{z}\de_{z} \zeta(z,x) + z \delta^{\mu}_{i}\gamma^{i} \de_{\mu}\zeta(z,x) -2\gamma^{z} \zeta(z,x) + m \zeta(z,x) = 0.
\eea
We can again use the Frobenius method near the boundary at $ z=0 $. We take the solution to be a series expansion in $ z $ of the form
\bea
\zeta(z,x)  = z^{\Delta} \sum_{l=0} c_{(l)}z^{l}.
\eea
Substituting this in \eqref{fermads} and from the leading order coefficient we get
\bea
\Delta = \begin{cases}
	2 - m ,\qquad &\text{with} \ \ \gamma^{z} c_{(0)}(x) = c_{(0)}(x)\\
	2 + m ,\qquad &\text{with} \ \ \gamma^{z} c_{(0)}(x) = -c_{(0)}(x).
\end{cases}
\eea
The most general solution can be written as, assuming $ m $ is positive,
\bea
\zeta(z,x) = c^{+}_{(0)}\,z^{2-m} +\dots + z^{2+m}\Big(c^{-}_{(0)} + c^{+}_{(2+m)} \log z \Big) + \dots
\eea
where the presence of the $ \log z $-term depends on the mass and is non-generic -- we will not need it in our discussions. 

The above discussion focusses on empty AdS background, where all the scalars are set to zero. This simplifies the discussion as the fermions and gravitino are all decoupled. We could in principle perform a similar analysis in the KS background, but the non-trivial couplings complicates the analysis substantially, and this is what we have done perturbatively in the main text.

\section{Supersymmetry of ${\cal N}=2$, $SU(2)\times SU(2)\times \mathbb{Z}_2$ truncation}\label{appd}
In this appendix we map the consistent truncation ansatz used in Liu-Szepietowski \cite{Liu:2011dw} (henceforth LS) to that of Cassani-Faedo \cite{Cassani} (henceforth CF). We then use this map to write down the fermionic SUSY variations in the notations of CF from which we then extract the BPS equations. We begin by defining the following one-forms
\begin{align}
\begin{split}
\sigma_1=c_{\psi/2}d\theta_1+s_{\psi/2}s_{\theta_1}d\phi_1~,~~~~&\Sigma_1=c_{\psi/2}d\theta_2+s_{\psi/2}s_{\theta_2}d\phi_2~,\\
\sigma_2=s_{\psi/2}d\theta_1-c_{\psi/2}s_{\theta_1}d\phi_1~,~~~~&\Sigma_2=s_{\psi/2}d\theta_2-c_{\psi/2}s_{\theta_2}d\phi_2~,\\
\sigma_3={1\over 2} d\psi+c_{\theta_1}d\phi_1~,~~~~&\Sigma_3={1\over 2} d\psi+c_{\theta_2}d\phi_2~.
\end{split}
\end{align}
where $ s_{\bullet} =\sin(\bullet) $ and $ c_{\bullet} =\cos(\bullet)$. These one-forms satisfy the $SU(2)\times SU(2)$ structure equations
\bea
d\sigma_i={1\over 2}\epsilon_{ijk}\sigma_i\wedge\sigma_j,~~~~~~~~d\Sigma_i={1\over 2}\epsilon_{ijk}\Sigma_i\wedge\Sigma_j~.
\eea
Using these forms we can endow a K\"{a}hler structure on $T^{1,1}$ as follows. We first define the following complex one-forms
\bea
\begin{split}
E_1={1\over\sqrt{6}}\(\sigma_1+i\sigma_2\)
\end{split}
\qquad  \qquad
\begin{split}
E_2={1\over\sqrt{6}}\(\Sigma_1+i\Sigma_2\).
\end{split}
\eea
Using these two complex one-forms we now define a basis of left-invariant forms on $T^{1,1}$ used in LS
\bea
J_{1} = \dfrac{i}{12} E_{1}\times \bar{E}_{1}~,\qquad J_{2} = \dfrac{i}{12} E_{2}\times \bar{E}_{2}~, \qquad \Omega = \dfrac{1}{6} E_{1} \times E_{2}~ , \qquad \eta = \dfrac{1}{3}\ee^{5}~,
\eea
where $\ee^5$ is defined in \eqref{oneforms1}. To compare with the notation of CF, we define $J_{\pm} = J_{1} \pm J_{2}$. The conversion now reads as follows
\bea
\eta_{\tx{LS}} = -\eta_{\tx{CF}}~,\qquad J_{+\, \tx{LS}} = -J_{\tx{CF}}~,\qquad  J_{-\, \tx{LS}} = -\Phi_{\tx{CF}}~,\qquad \Omega_{\tx{LS}}=\Omega_{\tx{CF}}~.
\eea
The LS metric is parametrized in the following way
\bea\label{LSzMetric}
ds_{LS}^{2} = e^{2A}ds_{5}^{2} + \dfrac{1}{6}e^{2B_{1}} E_{1}\bar{E}_{1} + \dfrac{1}{6}e^{2B_{2}} \hat{E}_{2}\hat{\bar{E}}_{2} + \dfrac{1}{9}e^{2C} (\eta+3A)^{2},
\eea
where $ \hat{E}_{2} = E_{2} + \alpha \bar{E}_{1}  $, $ \alpha $ being a complex scalar. In order to compare with the CF metric
\bea\label{CFMetric}
ds_{\tx{CF}}^{2} &=& e^{-\frac{8u-2v}{3}} ds_{5}^{2} + \dfrac{e^{2u}}{6} \cosh t \Big(e^{2w}(e_{1}^{2}+e_{2}^{2}) + e^{-2w}(e_{3}^{2}+e_{4}^{2})\Big) + \dfrac{e^{2v}}{9} (\eta+ 3A)^{2} \non\\
&& + \dfrac{e^{2u}}{3}\sinh t\Big( \cos\theta (e_{1}e_{3}+e_{2}e_{4}) + \sin\theta ( e_{1}e_{4} - e_{2}e_{3})\Big)~,
\eea
\eqref{LSzMetric} can be expanded in terms of $ \ee^{i} $'s defined in \eqref{oneforms1}. Upon comparing we obtain
\bea
A_{\tx{LS}} = -\dfrac{4u+v}{3} \bigg|_{\tx{CF}},\qquad \alpha_{\tx{LS}} = e^{2w} \tanh t ~e^{i\theta}|_{\tx{CF}} ,\qquad C_{\tx{LS}} = v |_{\tx{CF}}, \non\\
B_{1} = u+w -\dfrac{1}{2}\log\cosh t, \qquad B_{1} = u - w +\dfrac{1}{2}\log\cosh t~.
\eea
Similarly, from the expansion ansatz of the two form potentials, we get
\bea
e^{1}_{0} = -\bphi~, \ \ j_{0}^{2} = q~, \ \ b^{1}_{0} = \dfrac{1}{2}\overline{b^{\Omega}}~, \ \ b^{2}_{0} = \dfrac{1}{2} \overline{c^{\Omega}}~.
\eea
In the above equation we have written down the map only for fields turned on in the Klebanov-Strassler solution. Other relevant relations are as follows
\begin{align}
\begin{split}
&h_1^1=-d\bphi,\ \ f_0^1= \frac32 i~ \overline{b^{\Omega}}, \ \ f_0^2= \frac32 i~ \overline{c^{\Omega}}~,\\
&f_1^1= \frac12  d\overline{b^{\Omega}}, \ \ f_1^2= \frac12  d\overline{c^{\Omega}}~,\\
&\hat{f}_1^1= \frac12  d\overline{b^{\Omega}} +\frac i2 \tanh t~ d\bphi,\ \ \hat{f}_1^2= \frac12  d\overline{c^{\Omega}}~,\\
&\hat{f}_0^1 = \frac32 i~ \overline{b^{\Omega}}, \ \ \hat{f}_0^2 = \frac32 i~ \overline{c^{\Omega}}-\frac i2 q \tanh t ~,\\
&\hat{\cal F}_1^1= \frac i2 \sinh t~  \( - d\boi +\tanh t~ d\bphi  \), \ \ \hat{\cal F}_1^2= \frac12 \tanh t ~ d \cor ~,\\
&\hat{\cal F}_0^1= \frac 32 \boi \tanh t, \ \ \hat{\cal F}_0^2= \frac i2 \sinh t \( 3 \cor -q \tanh t\) ~.
\end{split}
\end{align}
One has to remember that $ \Re[b^{\Omega}]  = \Im[c^{\Omega}] = 0$. However, since many computations involve taking absolute values or the real and imaginary parts of products of functions, it is better to set this condition after making sure all such functions have been evaluated. Or one could set it and then be careful not to miss the $ i $ coming from $ b^{\Omega} = i\boi $. 
From the five form we get
\bea
-\dfrac{1}{2}(4+ \phi_{0}) = (k -q\bphi + 3 \boi\cor),
\eea
The notation for the axio-dilaton is $\tau=\tau_1+i\tau_2=C_{0}+i e^{-\phi}$. It will be convenient to write down the $SL(2,\mathbb{R})$ vielbein 
\bea
v_{1} = -(C_0e^{\phi/2}+i e^{-\phi/2})~,\qquad v_{2} = e^{\phi/2}~,
\eea
that appears explicitly in the SUSY variations of the fermions.
\subsection{SUSY variation of Fermions}\label{appd1}
\begin{subequations}\label{fermionSUSYvar}
\bea
\delta \zeta^{1} &= & \bigg[-\dfrac{i}{2\sqrt{2}}\Gamma.\de\phi \bigg]\epsilon -\dfrac{i}{2\sqrt{2}}e^{-2u-\frac{\phi}{2}}\Bigg[\Gamma\cdot\Big(e^{\phi}\de\cor + \cosh t \de \boi - \sinh t \de\bphi\Big) \non\\
&&\hspace{3cm} +e^{-\frac{4}{3}(u+v)}\Big(3\boi+3e^{\phi}\cosh t  \cor -qe^{\phi} \sinh t\Big)\bigg] \epsilon^{c}\\
\delta \zeta^{2} &=& -\dfrac{i}{2}e^{-2u-\frac{\phi}{2}}\cosh t\bigg[\Gamma\cdot\Big(\de \bphi -\tanh t \de \boi\Big) - e^{-\frac{4}{3}(u+v)+\phi}(3\tanh t \cor -q)\bigg]\epsilon\qquad\non\\
&&+ \frac i2 \bigg[\Gamma\cdot\de t + 3 \sinh t e^{-\frac{4}{3}(u+v)}\bigg]\epsilon^c\\
\delta \zeta^{3} &=& 2\sqrt{2}\bigg[-\dfrac{i}{2} \Gamma\cdot\de u -\dfrac{i}{2}e^{-\frac{2}{3}(5u-v)} +\dfrac{i}{8} e^{-\frac{4}{3}(4u+v)} (4+\phi_{0})\bigg] \epsilon \non\\
&& +\dfrac{i}{2\sqrt{2}}e^{-2u-\frac{\phi}{2}}\bigg[ \Gamma\cdot\Big(e^{\phi} \de\cor-\cosh t\de \boi + \sinh t \de\bphi  \Big) \non\\
&& \qquad\qquad +e^{-\frac{4}{3}(u+v)} \Big(3  \boi-3e^{\phi} \cosh t \cor + q e^{\phi}\sinh t\Big) \bigg] \epsilon^{c}\\
\delta \lambda^{u_3} &=& -\bigg[ -\dfrac{i}{6} \Gamma\cdot\de(u+v) +\dfrac{i}{6} e^{-\frac{2}{3}(5u-v)} -\dfrac{i}{2}\cosh t e^{-\frac{4}{3}(u+v)} +\dfrac{i}{12}e^{-\frac{4}{3}(4u+v)}(4+\phi_{0})\bigg]\epsilon\non\\
&&-\frac{i}{6} e^{-\frac{2}{3}(5u+2v) -\frac{\phi}{2}}\Big(3\boi -3e^{\phi}\cosh t\cor +q e^{\phi} \sinh t  \Big)\epsilon^{c}\\
\delta \Psi_{\mu} &=& \bigg[D_{\mu}+\dfrac{1}{6}\Gamma_{\mu}\calw \bigg]\epsilon +\bigg[\dfrac{\Gamma_{\mu}}{6}e^{-\frac{2}{3}(5u+2v)-\frac{\phi}{2}}\Big(3\boi -3e^{\phi}\cosh t \cor+ q e^{\phi}\sinh t\Big)\non\\
&&\qquad\qquad +\dfrac{1}{2}e^{-2u-\frac{\phi}{2}}\Big(\cosh t \de_{\mu}\boi - e^{\phi} \de_{\mu} \cor - \sinh t \de_{\mu}\bphi \Big)\bigg]\epsilon^{c}
\eea
\end{subequations}
The above SUSY transformation are taken from Eq. (102) of \cite{Halmagyi:2011yd} with the following definitions
\begin{align}
\begin{split}
\zeta^1_{\tx{here}}= \frac{1}{\sqrt{2}}\zeta^1_{\tx{there}}~,~~~~\zeta^2_{\tx{here}}= -\(\zeta^2_{\tx{there}}\)^c~,~~~~\zeta^3_{\tx{here}}= 2\sqrt{2}\zeta^3_{\tx{there}}~,~~~~\lambda^{u_3}_{\tx{here}}= -\xi^1_{\tx{there}}~.
\end{split}
\end{align}
The above field redefinitions are needed to extract the correctly normalized vielbeins of the scalar manifold such that they give rise to the metric ${\cal G}_{IJ}$ in \eqref{scalmetric}. The full scalar manifold can be seen as a direct product ${\cal Q} \otimes {\cal S}$ where ${\cal S}$ is a one dimensional very especial manifold and ${\cal Q}$ is twelve (real) dimensional quaternionic K\"ahler manifold.

Upon comparing with the notation of \cite{Ceresole:2000jd} one can extract the vielbeins and the SUSY variations of the scalars fields. In what follows we report this supergravity data. In writing down \eqref{fermionSUSYvar}, we have fixed some typos in \cite{Halmagyi:2011yd} which do not affect the BPS equations but do affect the metric on the scalar manifold .
\subsection{SUSY variation of Bosons}
The generic form of the SUSY variation of hyperino and gaugino in matter coupled ${\cal N}=2, D=5$ gauged supergravity is \cite{Ceresole:2000jd}
\bea\label{genferSUSY}
\begin{split}
\delta \zeta^A=-\frac i2 f^A_{iX} \slashed{\de}q^X \e^i+...\\
\delta \lambda^{x}_i=-\frac i2  \slashed{\de}\phi^x \e_i+...
\end{split}
\eea
where the dots denote the terms proportional to the gauging. All the fermions in the above formula are in the Symplectic-Majorana representation\footnote{In 5 dimensions, the minimal spinor is Dirac, so one cannot define a reality condition by relating the two minimal Weyl representations as in 4 dimensions. Instead, one takes two copies of Dirac to impose a complex conjugation condition relating them. The result is called a symplectic Majorana spinor. }. In the above formulas the index $i$ transforms in the fundamental representation of $SU(2)_R$ R-symmetry group, the index $A$ transforms in the fundamental representation of $USp(2n)$ (where $n$ is the number of hypermultiplets which in our case is three)\footnote{In \cite{Bertolini}, the fermionic sector was written in a sigma model form. The index carried by the fermions were treated on a similar footing as those of the scalars. While this notation allowed to write the fermionic Lagrangian and supersymmetry transformations (in the $U(1)$ truncation) compactly in terms of geometric quantities on the scalar manifold, it is not suitable for studying supersymmetry of the theory. It is not clear if a sigma model-type notation can be used for writing down the fermionic sector of the entire $\Z_2$ truncation.}.  The index $X$ labels coordinates on ${\cal Q}$ and the index $x$ labels coordinates on ${\cal S}$. To extract the vielbeins $f^A_{iX}$ on the quaternionic manifold, we first write down the symplectic-Majorana conditions for the fermions $\zeta^A$ that appears in the three hypermultiplets (here $A=1,2,3,4,5,6$) 
\bea\label{SMtoD}
\zeta^4=\(\zeta^1\)^c~,~~~~\zeta^5=\(\zeta^2\)^c~,~~~~\zeta^6=\(\zeta^3\)^c~.
\eea  
Here $\zeta^1,\zeta^2, \zeta^3$ are Dirac fermions that appear in \eqref{fermionSUSYvar}. The charge conjugation operation is defined as
\bea
\psi^c=\gamma_0C \psi^*~,
\eea
where $C$ is the charge conjugation matrix that satisfies the following properties
\bea
\begin{split}
&&C=-C^\dagger=-C^T=-C^{-1}=C^*~,\\
&&C^{-1}\gamma_\mu C =\gamma_\mu^T~.
\end{split}
\eea
Equation \eqref{genferSUSY} and \eqref{SMtoD} together imply the following relation between the vielbeins 
\begin{align}
\begin{split}
f^1_{1X}=f^4_{2X}~,~~~~f^1_{2X}=-f^4_{1X}~,\\
f^2_{1X}=f^5_{2X}~,~~~~f^2_{2X}=-f^5_{1X}~,\\
f^3_{1X}=f^6_{2X}~,~~~~f^3_{2X}=-f^6_{1X}~.
\end{split}
\end{align}
Upon comparing \eqref{genferSUSY} with \eqref{fermionSUSYvar} we get the following non-vanishing vielbeins of the quaternionic manifold ${\cal Q}$
\begin{align}\label{scalViel}
\begin{split}
&f^1_{1\phi}=\frac{1}{\sqrt{2}}~,~~~~f^1_{2\cor}=\frac{1}{\sqrt{2}}e^{-2u+\frac{\phi}{2}}~,\\
&f^1_{2\boi}=\frac{1}{\sqrt{2}}e^{-2u-\frac{\phi}{2}}\cosh t~,~~~~f^1_{2\bphi}=-\frac{1}{\sqrt{2}}e^{-2u-\frac{\phi}{2}}\sinh t~,\\
&f^2_{1\bphi}=e^{-2u-\frac{\phi}{2}}\cosh t~,~~~~f^2_{1\boi}=-e^{-2u-\frac{\phi}{2}}\sinh t~,~~~~f^2_{2t}=-1~,\\
&f^3_{1u}=2\sqrt{2}~,~~~~f^3_{2\cor}=-\frac{1}{\sqrt{2}}e^{-2u+\frac{\phi}{2}}~,\\
&f^3_{2\boi}=\frac{1}{\sqrt{2}}e^{-2u-\frac{\phi}{2}}\cosh t~,~~~~f^3_{2\bphi}=-\frac{1}{\sqrt{2}}e^{-2u-\frac{\phi}{2}}\sinh t~.
\end{split}
\end{align}
As a check of this result one can verify that with \eqref{scalViel}, one indeed reproduces the quaternionic metric in \eqref{scalmetric} via the following relation \cite{Ceresole:2001wi}
\bea\label{quatMetric}
g_{XY}\equiv C_{AB}\varepsilon^{ij} f^A_{iX}f^B_{jY}= f^{iA}_Xf_{YiA}~,
\eea
where $C_{AB}$ is the $USp(6)$ invariant tensor which in our convention reads
\begin{equation}C_{AB}=
\left(
\begin{array}{cc}
 0 & \mathbb{I}_3 \\
 -\mathbb{I}_3 & 0 \\
\end{array}
\right)~.
\end{equation}
In making this check we have to keep in mind the the metric $g_{XY}$ in \cite{Ceresole:2000jd} is defined upto a factor or 2 (see their Eq. (5.1)).

 The metric on the very special manifold, ${\cal S}$, parametrized by the scalar $u_3=-\frac13(u+v)$ in the vector multiplet, can be obtained by the following relations \cite{Halmagyi:2011yd}
\begin{align}\label{vsmMetric}
\begin{split}
G_{{\ci}{\cj}}&=X_\ci X_\cj-C_{\ci\cj\ck}X^\ck~,\\
X_\ci &=\frac12C_{\ci\cj\ck}X^\cj X^\ck~,\\
g_{xy}&= \de_x X^\ci \de_y X^\cj G_{\ci\cj}~,
\end{split}
\end{align}
where $X^\ci(\phi^x)$ are the embedding coordinates of the very special manifold that satisfies the following constraint
\bea
\frac 16 c_{\ci\cj\ck} X^\ci X^\cj X^\ck =1~.
\eea
For the supergravity model under consideration we have
\bea
X^0=e^{4u_3}~,~~~~X^1=e^{-2u_3}~,~~~~C_{011}=2~,
\eea
which, using \eqref{vsmMetric}, gives $g_{u+v,u+v}=\frac83$. Combining with $g_{uu}$ from \eqref{quatMetric}, we recover the ${\cal G}_{uu},{\cal G}_{vv},{\cal G}_{uv}$ components in \eqref{scalmetric}

We are now in a position to write down the bosonic SUSY variations. From \cite{Ceresole:2000jd} the generic form of the SUSY variation of the scalars in the hyper and vector multiplet is
\begin{align}\label{genbosSUSY}
\begin{split}
\delta q^X&=-i\bar{\epsilon}^i \zeta^A f_{iA}^X~,\\
\delta \phi^{x}&=\frac i2 \bar{\e}^i\lambda_i^x ~.
\end{split}
\end{align}
We remark that the index $X$ in $f^A_{iX}$ is raised an lowered by the metric $g_{XY}$ in \eqref{quatMetric} and not the metric in \eqref{scalmetric} which differs by a factor of two.
Using the vielbeins \eqref{scalViel} we find the following SUSY variations for the bosonic fields
\begin{subequations}\label{bosonSUSYvar}
\begin{align}
\delta e^a_\mu & = \frac12 \(\bar{\e} \gamma^a \Psi_\mu-\bar{\Psi}_\mu \gamma^a \e \)~,\\
\delta\phi &=\frac{i}{\sqrt{2}}\bar{\e}\zeta^1-\frac{i}{\sqrt{2}}\bar{\zeta^1}\e~,\\
\delta \bphi &=-\frac{i}{2}e^{2u+\frac{\phi}{2}} \[\cosh t\(\overline{\zeta^2}\e-\bar{\e}\zeta^2\)+\sinh t \(\overline{\chi_+}\e-\bar{\e}\chi_+\)\]~,\\
\delta \boi &=-\frac{i}{2}e^{2u+\frac{\phi}{2}} \[-\sinh t\(\overline{\zeta^2}\e-\bar{\e}\zeta^2\)+\cosh t \(\overline{\chi_+}\e-\bar{\e}\chi_+\)\]~,\\
\delta \cor &=-\frac i2 e^{2u-\frac{\phi}{2}}\big[\bar{\e}\chi_- -\overline{\chi_-}\e\big]~,\\
\delta t&= -\frac i2 \[\overline{\e^c}\zeta^2-\bar{\e}\(\zeta^2\)^c\]~, \\
\delta u & =-\frac{i}{4\sqrt{2}}\[\overline{\zeta^3}\e-\bar{\e}\zeta^3\]~, \\
\delta(u+v)&=-\frac{3i}{2}\[\bar{\e}\lambda^{u_3}-\overline{\lambda^{u_3}}\e\]~.
\end{align}
\end{subequations}
In the above equations we have used a new spinor $\chi_\pm$ which is defined as follows
\bea
\chi_\pm=-\frac{1}{\sqrt{2}}\(\zeta^4\pm\zeta^6\)~,~~~~\chi_\pm^c=\frac{1}{\sqrt{2}}\(\zeta^1\pm\zeta^3\)~.
\eea
These bosonic variations reduce to those of the $U(1)$ consistent truncation of \cite{Bertolini} upon using the following identification
\bea\label{bertolinimap}
\zeta_\phi=\sqrt{2}\zeta^1~,~~~~\zeta_b= e^{2u+\frac{\phi}{2}} \zeta^2~,~~~~\zeta_U= \frac{3}{2\sqrt{2}}\zeta^3-3\lambda^{u_3}~,~~~~\zeta_V= \frac{1}{\sqrt{2}}\zeta^3+3\lambda^{u_3}~.
\eea
This is the basis that we use in the main text.
\subsection{BPS equations from the fermionic variations}\label{appd3}
In this section, we extract from the fermionic variations, the BPS equations for flat domain walls where the metric takes the form \eqref{ksmetric} and all the scalars are function of the radial coordinate $\tau$.  The BPS equations take the form of a gradient flow \eqref{bpsu2} in terms of the superpotential $\cal W$ in \eqref{superpotential}. On supersymmetric configurations, $\delta \tx{(fermions)}=0$. We begin by splitting the SUSY variation parameter $\e$ appearing in \eqref{fermionSUSYvar} as follows:
\bea
\e=\e_+ +\e_-~,
\eea
with the property that 
\bea
\g ^5 \e_\pm =\pm \e_\pm~,~~~~\g ^5 \e_\pm^c =\mp \e_\pm^c~.
\eea
The $\g^5$ above the tangent space gamma matrix and is related to the curved space gamma matrix by $\g^a= e^a_{~\m}\G^\m$, where $e^a_{~\m}$ are the vielbeins of the five-dimensional spacetime. From the gauge fixed form of the metric in \eqref{ksmetric} we read that
\bea
\g^5= e^X \G^\tau~.
\eea
We now construct the projector
\bea
P_\pm =\frac12 \(1\pm \g^5\)~,
\eea
which satisfies
\bea
P_\pm^2=P_\pm~,~~~~P_+P_-=P_-P_+=0~.
\eea
Therefore we can write 
\bea
P_\pm \e=\e_{\pm}~,~~~~P_\pm \e^c=\e_{\mp}^c~.
\eea
Setting either $\e_+$ or $\e_-$ to zero kills half of the supersymmetries because $P_\pm$ is a half rank matrix. The choice is arbitrary and we choose to set $\e_+=\e_+^c=0$. Hitting the fermionic SUSY variations in \eqref{fermionSUSYvar} by  $P_\pm$ we extract the BPS equations. The system simplifies considerably if we start with the variation of $\lambda^{u_3}$. From the term proportional to $\e^c$ we find a constraint
\bea\label{constraint1}
3\boi -3e^{\phi}\cosh t\cor +q e^{\phi} \sinh t=0~.
\eea
Next we move to the $\tau$ component of the gravitino variation. Since we have guage fixed $\Psi_\tau=0$, we find upon using \eqref{constraint1} another equation from the term proportional to $\e^c$ 
\bea\label{constraint2}
\cosh t \de_{\tau}\boi - e^{\phi} \de_{\tau} \cor - \sinh t \de_{\tau}\bphi =0~.
\eea
Using \eqref{constraint1} and \eqref{constraint2} we see that in variation of $\zeta^3$ the entire piece proportional to $\e^c$ vanish. From the remaining equations we get
\begin{subequations}
\begin{align}
&\de_\tau \f=0~,\\
&e^{-X}\Big(e^{\phi}\de_\tau\cor + \cosh t \de_\tau \boi - \sinh t \de_\tau\bphi\Big) +e^{-\frac{4}{3}(u+v)}\Big(3\boi+3e^{\phi}\cosh t  \cor -qe^{\phi} \sinh t\Big)=0~,\\
&e^{-X}\Big(\de_\tau \bphi -\tanh t \de_\tau \boi\Big) + e^{-\frac{4}{3}(u+v)+\phi}(3\tanh t \cor -q)=0~,\\
&e^{-X}\de_\tau t + 3 \sinh t e^{-\frac{4}{3}(u+v)}=0~,\\
&\dfrac{1}{2} e^ {-X}\de_\tau u -\dfrac{1}{2}e^{-\frac{2}{3}(5u-v)} +\dfrac{1}{8} e^{-\frac{4}{3}(4u+v)} (4+\phi_{0})=0~,\\
&\dfrac{1}{6} e^{-X}\de_{\tau}(u+v) +\dfrac{1}{6} e^{-\frac{2}{3}(5u-v)} -\dfrac{1}{2}\cosh t e^{-\frac{4}{3}(u+v)} +\dfrac{1}{12}e^{-\frac{4}{3}(4u+v)}(4+\phi_{0})=0~.
\end{align}
\end{subequations}
Upon using the constraints \eqref{constraint1} and \eqref{constraint2} it is straightforward to show that these equation reduce to \eqref{bpsu2}.

\bibliography{ref}

\providecommand{\href}[2]{#2}\begingroup\raggedright\begin{thebibliography}{10}

\bibitem{Willy}
W.~Fischler, A.~Kashani-Poor, R.~McNees, and S.~Paban, ``{The Acceleration of
  the universe, a challenge for string theory},'' {\em JHEP} {\bf 07} (2001)
  003, \href{http://xxx.lanl.gov/abs/hep-th/0104181}{{\tt hep-th/0104181}}.

\bibitem{ISS}
K.~A. Intriligator, N.~Seiberg, and D.~Shih, ``{Dynamical SUSY breaking in
  meta-stable vacua},'' {\em JHEP} {\bf 04} (2006) 021,
  \href{http://xxx.lanl.gov/abs/hep-th/0602239}{{\tt hep-th/0602239}}.

\bibitem{Argurio:2006ny}
R.~Argurio, M.~Bertolini, S.~Franco, and S.~Kachru, ``{Gauge/gravity duality
  and meta-stable dynamical supersymmetry breaking},'' {\em JHEP} {\bf 01}
  (2007) 083, \href{http://xxx.lanl.gov/abs/hep-th/0610212}{{\tt
  hep-th/0610212}}.

\bibitem{Argurio:2007qk}
R.~Argurio, M.~Bertolini, S.~Franco, and S.~Kachru, ``{Meta-stable vacua and
  D-branes at the conifold},'' {\em JHEP} {\bf 06} (2007) 017,
  \href{http://xxx.lanl.gov/abs/hep-th/0703236}{{\tt hep-th/0703236}}.

\bibitem{WillyVadimCK}
W.~Fischler, V.~Kaplunovsky, C.~Krishnan, L.~Mannelli, and M.~A.~C. Torres,
  ``{Meta-Stable Supersymmetry Breaking in a Cooling Universe},'' {\em JHEP}
  {\bf 03} (2007) 107, \href{http://xxx.lanl.gov/abs/hep-th/0611018}{{\tt
  hep-th/0611018}}.

\bibitem{KKLT}
S.~Kachru, R.~Kallosh, A.~D. Linde, and S.~P. Trivedi, ``{De Sitter vacua in
  string theory},'' {\em Phys. Rev.} {\bf D68} (2003) 046005,
  \href{http://xxx.lanl.gov/abs/hep-th/0301240}{{\tt hep-th/0301240}}.

\bibitem{GiddingsKP}
S.~B. Giddings, S.~Kachru, and J.~Polchinski, ``{Hierarchies from fluxes in
  string compactifications},'' {\em Phys. Rev.} {\bf D66} (2002) 106006,
  \href{http://xxx.lanl.gov/abs/hep-th/0105097}{{\tt hep-th/0105097}}.

\bibitem{Sethi}
S.~Sethi, ``{Supersymmetry Breaking by Fluxes},''
  \href{http://xxx.lanl.gov/abs/1709.03554}{{\tt 1709.03554}}.

\bibitem{KPV}
S.~Kachru, J.~Pearson, and H.~L. Verlinde, ``{Brane / flux annihilation and the
  string dual of a nonsupersymmetric field theory},'' {\em JHEP} {\bf 06}
  (2002) 021, \href{http://xxx.lanl.gov/abs/hep-th/0112197}{{\tt
  hep-th/0112197}}.

\bibitem{Jarah}
J.~Evslin, C.~Krishnan, and S.~Kuperstein, ``{Cascading quivers from decaying
  D-branes},'' {\em JHEP} {\bf 08} (2007) 020,
  \href{http://xxx.lanl.gov/abs/0704.3484}{{\tt 0704.3484}}.

\bibitem{KS}
I.~R. Klebanov and M.~J. Strassler, ``{Supergravity and a confining gauge
  theory: Duality cascades and chi SB resolution of naked singularities},''
  {\em JHEP} {\bf 08} (2000) 052,
  \href{http://xxx.lanl.gov/abs/hep-th/0007191}{{\tt hep-th/0007191}}.

\bibitem{KW}
I.~R. Klebanov and E.~Witten, ``{Superconformal field theory on three-branes at
  a Calabi-Yau singularity},'' {\em Nucl. Phys.} {\bf B536} (1998) 199--218,
  \href{http://xxx.lanl.gov/abs/hep-th/9807080}{{\tt hep-th/9807080}}.

\bibitem{DKM}
O.~DeWolfe, S.~Kachru, and M.~Mulligan, ``{A Gravity Dual of Metastable
  Dynamical Supersymmetry Breaking},'' {\em Phys. Rev.} {\bf D77} (2008)
  065011, \href{http://xxx.lanl.gov/abs/0801.1520}{{\tt 0801.1520}}.

\bibitem{Bertolini}
M.~Bertolini, D.~Musso, I.~Papadimitriou, and H.~Raj, ``{A goldstino at the
  bottom of the cascade},'' {\em JHEP} {\bf 11} (2015) 184,
  \href{http://xxx.lanl.gov/abs/1509.03594}{{\tt 1509.03594}}.

\bibitem{Aharony}
O.~Aharony, A.~Buchel, and A.~Yarom, ``{Holographic renormalization of
  cascading gauge theories},'' {\em Phys. Rev.} {\bf D72} (2005) 066003,
  \href{http://xxx.lanl.gov/abs/hep-th/0506002}{{\tt hep-th/0506002}}.

\bibitem{KT}
I.~R. Klebanov and A.~A. Tseytlin, ``{Gravity duals of supersymmetric SU(N) x
  SU(N+M) gauge theories},'' {\em Nucl. Phys.} {\bf B578} (2000) 123--138,
  \href{http://xxx.lanl.gov/abs/hep-th/0002159}{{\tt hep-th/0002159}}.

\bibitem{Argurio}
R.~Argurio, M.~Bertolini, D.~Musso, F.~Porri, and D.~Redigolo, ``{Holographic
  Goldstino},'' {\em Phys. Rev.} {\bf D91} (2015), no.~12 126016,
  \href{http://xxx.lanl.gov/abs/1412.6499}{{\tt 1412.6499}}.

\bibitem{Buchel}
A.~Buchel, ``{Effective Action of the Baryonic Branch in String Theory Flux
  Throats},'' {\em JHEP} {\bf 09} (2014) 117,
  \href{http://xxx.lanl.gov/abs/1405.1518}{{\tt 1405.1518}}.

\bibitem{Cassani}
D.~Cassani and A.~F. Faedo, ``{A Supersymmetric consistent truncation for
  conifold solutions},'' {\em Nucl. Phys.} {\bf B843} (2011) 455--484,
  \href{http://xxx.lanl.gov/abs/1008.0883}{{\tt 1008.0883}}.

\bibitem{Bena:2010pr}
I.~Bena, G.~Giecold, M.~Grana, N.~Halmagyi, and F.~Orsi, ``{Supersymmetric
  Consistent Truncations of IIB on $T^{1,1}$},'' {\em JHEP} {\bf 04} (2011)
  021, \href{http://xxx.lanl.gov/abs/1008.0983}{{\tt 1008.0983}}.

\bibitem{Halmagyi:2011yd}
N.~Halmagyi, J.~T. Liu, and P.~Szepietowski, ``{On N = 2 Truncations of IIB on
  $ T^{1,1} $},'' {\em JHEP} {\bf 07} (2012) 098,
  \href{http://xxx.lanl.gov/abs/1111.6567}{{\tt 1111.6567}}.

\bibitem{Liu:2011dw}
J.~T. Liu and P.~Szepietowski, ``{Supersymmetry of consistent massive
  truncations of IIB supergravity},'' {\em Phys. Rev.} {\bf D85} (2012) 126010,
  \href{http://xxx.lanl.gov/abs/1103.0029}{{\tt 1103.0029}}.

\bibitem{PT}
G.~Papadopoulos and A.~A. Tseytlin, ``{Complex geometry of conifolds and
  five-brane wrapped on two sphere},'' {\em Class. Quant. Grav.} {\bf 18}
  (2001) 1333--1354, \href{http://xxx.lanl.gov/abs/hep-th/0012034}{{\tt
  hep-th/0012034}}.

\bibitem{Ashmore:2016oug}
A.~Ashmore, M.~Gabella, M.~Graña, M.~Petrini, and D.~Waldram, ``{Exactly
  marginal deformations from exceptional generalised geometry},'' {\em JHEP}
  {\bf 01} (2017) 124, \href{http://xxx.lanl.gov/abs/1605.05730}{{\tt
  1605.05730}}.

\bibitem{Gubser:2000nd}
S.~S. Gubser, ``{Curvature singularities: The Good, the bad, and the naked},''
  {\em Adv. Theor. Math. Phys.} {\bf 4} (2000) 679--745,
  \href{http://xxx.lanl.gov/abs/hep-th/0002160}{{\tt hep-th/0002160}}.

\bibitem{Herzog:2002ih}
C.~P. Herzog, I.~R. Klebanov, and P.~Ouyang, ``{D-branes on the conifold and
  N=1 gauge / gravity dualities},'' in {\em {Progress in string, field and
  particle theory: Proceedings, NATO Advanced Study Institute, EC Summer
  School, Cargese, France, June 25-July 11, 2002}}, pp.~189--223, 2002.
\newblock \href{http://xxx.lanl.gov/abs/hep-th/0205100}{{\tt hep-th/0205100}}.
\newblock [,383(2002)].

\bibitem{GranaMinasian}
A.~Butti, M.~Grana, R.~Minasian, M.~Petrini, and A.~Zaffaroni, ``{The Baryonic
  branch of Klebanov-Strassler solution: A supersymmetric family of SU(3)
  structure backgrounds},'' {\em JHEP} {\bf 03} (2005) 069,
  \href{http://xxx.lanl.gov/abs/hep-th/0412187}{{\tt hep-th/0412187}}.

\bibitem{CKK2}
C.~Krishnan and S.~Kuperstein, ``{The Mesonic Branch of the Deformed
  Conifold},'' {\em JHEP} {\bf 05} (2008) 072,
  \href{http://xxx.lanl.gov/abs/0802.3674}{{\tt 0802.3674}}.

\bibitem{Kuperstein}
S.~Kuperstein, B.~Truijen, and T.~Van~Riet, ``{Non-SUSY fractional branes},''
  {\em JHEP} {\bf 03} (2015) 161, \href{http://xxx.lanl.gov/abs/1411.3358}{{\tt
  1411.3358}}.

\bibitem{BGGHM}
I.~Bena, G.~Giecold, M.~Grana, N.~Halmagyi, and S.~Massai, ``{The backreaction
  of anti-D3 branes on the Klebanov-Strassler geometry},'' {\em JHEP} {\bf 06}
  (2013) 060, \href{http://xxx.lanl.gov/abs/1106.6165}{{\tt 1106.6165}}.

\bibitem{Loewy:2001pq}
A.~Loewy and J.~Sonnenschein, ``{On the holographic duals of N=1 gauge
  dynamics},'' {\em JHEP} {\bf 08} (2001) 007,
  \href{http://xxx.lanl.gov/abs/hep-th/0103163}{{\tt hep-th/0103163}}.

\bibitem{Kuperstein:2003yt}
S.~Kuperstein and J.~Sonnenschein, ``{Analytic nonsupersymmtric background dual
  of a confining gauge theory and the corresponding plane wave theory of
  hadrons},'' {\em JHEP} {\bf 02} (2004) 015,
  \href{http://xxx.lanl.gov/abs/hep-th/0309011}{{\tt hep-th/0309011}}.

\bibitem{Henneaux}
M.~Henneaux, ``{Boundary terms in the AdS / CFT correspondence for spinor
  fields},'' in {\em {Mathematical methods in modern theoretical physics.
  Proceedings, International Meeting, School and Workshop, ISPM'98, Tbilisi,
  Georgia, September 5-18, 1998}}, pp.~161--170, 1998.
\newblock \href{http://xxx.lanl.gov/abs/hep-th/9902137}{{\tt hep-th/9902137}}.

\bibitem{Papadimitriou:2017kzw}
I.~Papadimitriou, ``{Supercurrent anomalies in 4d SCFTs},'' {\em JHEP} {\bf 07}
  (2017) 038, \href{http://xxx.lanl.gov/abs/1703.04299}{{\tt 1703.04299}}.

\bibitem{An:2017ihs}
O.~S. An, ``{Anomaly-corrected supersymmetry algebra and supersymmetric
  holographic renormalization},'' {\em JHEP} {\bf 12} (2017) 107,
  \href{http://xxx.lanl.gov/abs/1703.09607}{{\tt 1703.09607}}.

\bibitem{Papadimitriou:2004rz}
I.~Papadimitriou and K.~Skenderis, ``{Correlation functions in holographic RG
  flows},'' {\em JHEP} {\bf 10} (2004) 075,
  \href{http://xxx.lanl.gov/abs/hep-th/0407071}{{\tt hep-th/0407071}}.

\bibitem{deHaro}
S.~de~Haro, S.~N. Solodukhin, and K.~Skenderis, ``{Holographic reconstruction
  of space-time and renormalization in the AdS / CFT correspondence},'' {\em
  Commun. Math. Phys.} {\bf 217} (2001) 595--622,
  \href{http://xxx.lanl.gov/abs/hep-th/0002230}{{\tt hep-th/0002230}}.

\bibitem{Papadimitriou:2004ap}
I.~Papadimitriou and K.~Skenderis, ``{AdS / CFT correspondence and geometry},''
  {\em IRMA Lect. Math. Theor. Phys.} {\bf 8} (2005) 73--101,
  \href{http://xxx.lanl.gov/abs/hep-th/0404176}{{\tt hep-th/0404176}}.

\bibitem{Avinash}
C.~Krishnan, A.~Raju, and P.~N.~B. Subramanian, ``{Dynamical boundary for
  anti–de Sitter space},'' {\em Phys. Rev.} {\bf D94} (2016), no.~12 126011,
  \href{http://xxx.lanl.gov/abs/1609.06300}{{\tt 1609.06300}}.

\bibitem{Bianchi}
M.~Bianchi, D.~Z. Freedman, and K.~Skenderis, ``{How to go with an RG flow},''
  {\em JHEP} {\bf 08} (2001) 041,
  \href{http://xxx.lanl.gov/abs/hep-th/0105276}{{\tt hep-th/0105276}}.

\bibitem{Argurio0}
R.~Argurio, M.~Bertolini, L.~Pietro, F.~Porri, and D.~Redigolo, ``{Supercurrent
  multiplet correlators at weak and strong coupling},'' {\em JHEP} {\bf 04}
  (2014) 123, \href{http://xxx.lanl.gov/abs/1310.6897}{{\tt 1310.6897}}.

\bibitem{Halmagyi}
I.~Bena, M.~Grana, and N.~Halmagyi, ``{On the Existence of Meta-stable Vacua in
  Klebanov-Strassler},'' {\em JHEP} {\bf 09} (2010) 087,
  \href{http://xxx.lanl.gov/abs/0912.3519}{{\tt 0912.3519}}.

\bibitem{Polchinski}
J.~Polchinski, ``{Brane/antibrane dynamics and KKLT stability},''
  \href{http://xxx.lanl.gov/abs/1509.05710}{{\tt 1509.05710}}.

\bibitem{Dasgupta}
E.~A. Bergshoeff, K.~Dasgupta, R.~Kallosh, A.~Van~Proeyen, and T.~Wrase, ``{$
  \overline{\mathrm{D}3} $ and dS},'' {\em JHEP} {\bf 05} (2015) 058,
  \href{http://xxx.lanl.gov/abs/1502.07627}{{\tt 1502.07627}}.

\bibitem{Wrase}
R.~Kallosh and T.~Wrase, ``{Emergence of Spontaneously Broken Supersymmetry on
  an Anti-D3-Brane in KKLT dS Vacua},'' {\em JHEP} {\bf 12} (2014) 117,
  \href{http://xxx.lanl.gov/abs/1411.1121}{{\tt 1411.1121}}.

\bibitem{Ceresole:2000jd}
A.~Ceresole and G.~Dall'Agata, ``{General matter coupled N=2, D = 5 gauged
  supergravity},'' {\em Nucl. Phys.} {\bf B585} (2000) 143--170,
  \href{http://xxx.lanl.gov/abs/hep-th/0004111}{{\tt hep-th/0004111}}.

\bibitem{Ceresole:2001wi}
A.~Ceresole, G.~Dall'Agata, R.~Kallosh, and A.~Van~Proeyen, ``{Hypermultiplets,
  domain walls and supersymmetric attractors},'' {\em Phys. Rev.} {\bf D64}
  (2001) 104006, \href{http://xxx.lanl.gov/abs/hep-th/0104056}{{\tt
  hep-th/0104056}}.

\end{thebibliography}\endgroup
\bibliographystyle{utphys}
\end{document}